\documentclass[twocolumn,twocolappendix]{aastex701}
\usepackage{threeparttablex}

\usepackage{booktabs}
\usepackage{amsmath}
\usepackage{natbib}
\usepackage{rotating}
\usepackage{tabularx}
\usepackage{multirow}

\begin{document}

\title{An extremely bright slow-rising afterglow from an off-axis jet in GRB 260310A}

\correspondingauthor{Yu-Han Yang, Eleonora Troja}
\author[0000-0003-0691-6688]{Yu-Han Yang}
\affiliation{Department of Physics, University of Rome ``Tor Vergata'', via della Ricerca Scientifica 1, I-00133 Rome, Italy}
\email[show]{yuhan.yang@roma2.infn.it}

\author[0000-0003-4631-1528]{Roberto Ricci}
\affiliation{Department of Physics, University of Rome ``Tor Vergata'', via della Ricerca Scientifica 1, I-00133 Rome, Italy}
\affiliation{INAF - Istituto di Radioastronomia, via Gobetti 101, I-40129 Bologna, Italy}
\email{}

\author[0000-0002-1869-7817]{Eleonora Troja}
\affiliation{Department of Physics, University of Rome ``Tor Vergata'', via della Ricerca Scientifica 1, I-00133 Rome, Italy}
\email[show]{eleonora.troja@uniroma2.it}

\author[0009-0004-9520-5822]{Muskan Yadav}
\affiliation{Department of Physics, University of Rome ``Tor Vergata'', via della Ricerca Scientifica 1, I-00133 Rome, Italy}
\email{muskan.yadav@students.uniroma2.eu}

\author[0000-0002-5596-5059]{Yi-Han Iris Yin}
\affiliation{The Hong Kong Institute for Astronomy and Astrophysics, The University of Hong Kong, Hong Kong, China}
\affiliation{Department of Physics, The University of Hong Kong, Pokfulam Road, Hong Kong, China}
\email{iris.yh.yin@connect.hku.hk}

\author[0000-0002-7158-5099]{Rub\'en S\'anchez-Ram\'irez}
\affiliation{Instituto de Astrof\'isica de Andaluc\'ia (IAA-CSIC), Glorieta de la Astronom\'ia s/n, 18008 Granada, Spain}
\email{ruben@iaa.es}

\author[0000-0002-9700-0036]{Brendan O'Connor}
\affiliation{McWilliams Center for Cosmology and Astrophysics, Department of Physics, Carnegie Mellon University, Pittsburgh, PA 15213, USA}
\email{boconno2@andrew.cmu.edu}

\author[0009-0007-6886-4082]{Niccol\'o Passaleva}
\affiliation{Department of Physics, University of Rome ``Sapienza'', P.le Aldo Moro 2, I-00185 Rome, Italy}
\affiliation{Department of Physics, University of Rome ``Tor Vergata'', via della Ricerca Scientifica 1, I-00133 Rome, Italy}
\email{niccolo.passaleva@uniroma1.it}

\author[0000-0003-2999-3563]{Alberto J. Castro-Tirado}
\affiliation{Instituto de Astrof\'isica de Andaluc\'ia (IAA-CSIC), Glorieta de la Astronom\'ia s/n, 18008 Granada, Spain}
\affiliation{Ingeniería de Sistemas y Autom\'atica, Universidad de M\'alaga, Unidad Asociada al CSIC por el IAA, Escuela de Ingenier\'ias Industriales, Arquitecto Francisco Pe\~nalosa, 6, Campanillas, 29071 M\'alaga, Spain}
\email{ajct@iaa.es}

\author[0000-0002-8680-8718]{Hendrik van Eerten}
\affiliation{Department of Physics, University of Bath, Building 3 West, Bath BA2 7AY, UK}
\email{}

\author[0000-0001-6849-1270]{Simone Dichiara}
\affiliation{Department of Astronomy and Astrophysics, The Pennsylvania State University, 525 Davey Lab, University Park, PA 16802, USA}
\email{sbd5667@psu.edu}

\author[0000-0003-1394-7044]{Vincenzo Galluzzi}
\affiliation{INAF - Istituto di Radioastronomia, via Gobetti 101, I-40129 Bologna, Italy}
\affiliation{INAF – Italian Centre for Astronomical Archives c/o Osservatorio Astronomico di Trieste, Via Tiepolo 11, I-34143, Trieste, Italy}
\email{vincenzo.galluzzi@inaf.it}

\author[0009-0009-7526-4522]{Narjes Shahamat Dehsorkh}
\affiliation{Department of Physics, University of Rome ``Tor Vergata'', via della Ricerca Scientifica 1, I-00133 Rome, Italy}
\email{}

\author[0000-0002-3777-6182]{Iv\'an Agudo}
\affiliation{Instituto de Astrof\'isica de Andaluc\'ia (IAA-CSIC), Glorieta de la Astronom\'ia s/n, 18008 Granada, Spain}
\email{iagudo@iaa.es}

\author[0000-0003-0487-1105]{Jes\'us Aceituno}
\affiliation{Instituto de Astrof\'isica de Andaluc\'ia (IAA-CSIC), Glorieta de la Astronom\'ia s/n, 18008 Granada, Spain}
\email{aceitun@iaa.es}

\author[0009-0001-0574-2332]{Malte Busmann}
    \affiliation{University Observatory, Faculty of Physics, Ludwig-Maximilians-Universität München, Scheinerstr. 1, 81679 Munich, Germany}
    \affiliation{Excellence Cluster ORIGINS, Boltzmannstr. 2, 85748 Garching, Germany}
    \email{m.busmann@physik.lmu.de}

\author[0000-0001-7920-4564]{Maria D. Caballero-Garc\'ia}
\affiliation{Instituto de Astrof\'isica de Andaluc\'ia (IAA-CSIC), Glorieta de la Astronom\'ia s/n, 18008 Granada, Spain}
\email{mcaballero@iaa.es}

\author[0009-0009-4604-9639]{Emilio Fern\'andez-Garc\'ia}
\affiliation{Instituto de Astrof\'isica de Andaluc\'ia (IAA-CSIC), Glorieta de la Astronom\'ia s/n, 18008 Granada, Spain}
\email{emifdez@iaa.es}

\author[0000-0003-3270-7644]{Daniel Gruen}
	\affiliation{University Observatory, Faculty of Physics, Ludwig-Maximilians-Universität München, Scheinerstr. 1, 81679 Munich, Germany}
	\affiliation{Excellence Cluster ORIGINS, Boltzmannstr. 2, 85748 Garching, Germany}
	\email{daniel.gruen@lmu.de}

\author[0000-0003-4268-6277]{Maria Gritsevich}
\affiliation{Instituto de Astrof\'isica de Andaluc\'ia (IAA-CSIC), Glorieta de la Astronom\'ia s/n, 18008 Granada, Spain}
\email{maria@iaa.es}

\author[0000-0003-2628-6468]{Sergiy Guziy}
\affiliation{Instituto de Astrof\'isica de Andaluc\'ia (IAA-CSIC), Glorieta de la Astronom\'ia s/n, 18008 Granada, Spain}
\affiliation{Petro Mohyla Black Sea National University, Mykolaiv 54000, Ukraine}
\email{gss@iaa.es}

\author[0000-0002-4711-7658]{David Hiriart}
\affiliation{Instituto de Astronom\'{\i}a, Universidad Nacional Aut\'onoma de M\'exico (IA-UNAM), carretera Tijuana-Ensenada, km. 107, C.P. 22860 Ensenada, Baja California, M\'exico}
\email{hiriart@astro.unam.mx}

\author[0000-0002-7400-4608]{You-Dong Hu}
\affiliation{Guangxi Key Laboratory for Relativistic Astrophysics, School of Physical Science and Technology, Guangxi University, Nanning 530004, China}
\email{huyoudong0772@hotmail.com}

\author[0000-0003-3922-7416]{Martin Jel\'inek}
\affiliation{Astronomical Institute, Czech Academy of Sciences, Fri\v{c}ova 298, 251 65 Ond\v{r}ejov, Czech Republic}
\email{mates@asu.cas.cz}

\author[0000-0002-2715-8460]{Alexander Kutyrev}
\affiliation{Astrophysics Science Division, NASA Goddard Space Flight Center, 8800 Greenbelt Rd, Greenbelt, MD 20771, USA}
\affiliation{Department of Astronomy, University of Maryland, College Park, MD 20742-4111, USA}
\email{alexander.s.kutyrev@nasa.gov}

\author[0000-0002-0636-9138]{Alzbeta Malenakova}
\affiliation{Astronomical Institute, Czech Academy of Sciences, Fri\v{c}ova 298, 251 65 Ond\v{r}ejov, Czech Republic}
\email{alzbeta.malenakova@email.cz}

\author[0009-0008-5788-3584]{Filip Novotny}
\affiliation{Astronomical Institute, Czech Academy of Sciences, Fri\v{c}ova 298, 251 65 Ond\v{r}ejov, Czech Republic}
\affiliation{Masaryk University Brno, Faculty of Natural Sciences, Kotl\'{a}\v{r}sk\'{a} 2, 602 00 Brno, Czech Republic}
\email{fnovotny@physics.muni.cz}

\author[0000-0002-7273-3671]{Ignacio P\'erez-Garc\'ia}
\affiliation{Instituto de Astrof\'isica de Andaluc\'ia (IAA-CSIC), Glorieta de la Astronom\'ia s/n, 18008 Granada, Spain}
\email{ipg@iaa.es}

\author[0000-0002-1474-1980]{Shashi B. Pandey}
\affiliation{Aryabhatta Research Institute of Observational Sciences (ARIES), Manora Peak, Nainital, Uttarakhand 263001, India}
\email{shashiaries0@gmail.com}

\author[0009-0009-4453-8260]{Jorma Ryske}
\affiliation{URSA Astronomical Association, Kopernikuksentie 1, 00130 Helsinki, Finland}
\email{jorma.ryske@gmail.com}

\author[0000-0002-9404-6952]{Alfredo Sota}
\affiliation{Instituto de Astrof\'isica de Andaluc\'ia (IAA-CSIC), Glorieta de la Astronom\'ia s/n, 18008 Granada, Spain}
\email{sota@iaa.es}

\author[0000-0002-4147-2878]{Jan Strobl}
\affiliation{Astronomical Institute, Czech Academy of Sciences, Fri\v{c}ova 298, 251 65 Ond\v{r}ejov, Czech Republic}
\email{jan.strobl@asu.cas.cz}

\author[0009-0001-2089-9899]{Hira Waseem}
\affiliation{Department of Physics, University of Rome ``Sapienza'', P.le Aldo Moro 2, I-00185 Rome, Italy}
\affiliation{Department of Physics, University of Rome ``Tor Vergata'', via della Ricerca Scientifica 1, I-00133 Rome, Italy}
\email{}

\author[0009-0004-7113-8258]{Siyu Wu}
\affiliation{Instituto de Astrof\'isica de Andaluc\'ia (IAA-CSIC), Glorieta de la Astronom\'ia s/n, 18008 Granada, Spain}
\email{wusiyu.11@outlook.com}

\begin{abstract}
We present a multi-wavelength study of GRB\,260310A, a nearby long-duration gamma-ray burst (GRB) at $z\simeq0.153$ associated with a broad-lined Type Ic supernova. Despite its modest prompt gamma-ray output, $E_{\gamma,\rm iso}\simeq3.5\times10^{50}$ erg,  GRB\,260310A exhibits one of the brightest afterglows ever observed in the X-ray, optical, and radio bands. 
Its apparent brightness is not its only remarkable feature. 
The optical afterglow displays a delayed onset, characterized by a slow rising phase, with slope $\alpha\approx-1$, and a late peak at $\approx$0.1 d. 
We argue that the combination of weak prompt emission, hard peak energy, and late afterglow onset is naturally explained by a GRB jet viewed off-axis.
The radio spectral energy distributions are consistent with synchrotron radiation and indicate the presence of both reverse- and forward-shock components, thus providing a first test of reverse-shock models in an off-axis geometry.
The X-ray afterglow displays a prominent rebrightening, monitored for up to $\approx$68 d with no evidence of spectral evolution. A low level of linear polarization, $\Pi\approx1.7\%$, is measured at 15 GHz at $T_0+55$ d and suggests that, at these late times, the forward-shock is the dominant emission component from radio to X-rays. 
This late-time rebrightening represents a critical test for the two-component jet model.  If interpreted as the emergence of a narrow jet core viewed further off-axis, it would imply extreme luminosities and energetics for an on-axis observer. 

\end{abstract}


\section{Introduction} 

Gamma-ray bursts (GRBs) are among the brightest and most energetic explosions in the Universe \citep{review_zhang_mesaros, review_piran, review_geherls}.  Detected as brief flashes of gamma-rays from cosmological distances, they are followed by a longer lasting afterglow visible across the electromagnetic spectrum, from X-rays to radio wavelengths \citep{costa97, VanParadijs1997, Frail1997}. 
Their extreme luminosities are powered by a central compact object - either an accreting black hole \citep{narayan1992} or a massive neutron star \citep{usov1992} - driving an ultra-relativistic jet \citep{Rhoads1999,Sari1999}. 

Most GRBs detected to date are those whose jets are closely aligned with our line of sight (on-axis), so that their emission is strongly Doppler-boosted and easily detectable.  As a result, on-axis observers probe only the inner and most energetic portion of the outflow (jet core), while remaining largely insensitive to the angular structure at wider angles \citep{granot2002, Rossi2002, Zhang2002}. Any stratification in energy and Lorentz factor outside the core, as well as the presence of a cocoon or slower wings \citep{RamirezRuiz2002,Nakar2017,Izzo2019}, is essentially hidden from view. In this sense, standard on-axis GRBs provide an incomplete picture of the jet, which can often be approximated as a simple “top-hat” even when the outflow is intrinsically structured.

However, the majority of GRB jets are oriented away from the observer \citep{Huang2002,granot2002,Ryan2015,Chakyar2025} and remain largely unseen. As the viewing angle increases, relativistic beaming reduces both the observed flux and the characteristic photon energies, causing these events to fall below the sensitivity thresholds of traditional gamma-ray detectors.
Thus, GRBs seen off-axis are expected to manifest as weak high-energy transients, or even gamma-ray dark events \citep{Rhoads1999,Yamazaki2003,Lazzati2017b,Xu2023,ZhengLu2026}, whose multi-wavelength afterglows evolve on longer timescales compared to classical on-axis bursts \citep{granot2002,vanEerten2013,Kathirgamaraju2018,Ryan2020}.
Uncovering this population of misaligned jets is essential to provide a complete census of relativistic explosions and constrain their jet structure, which encodes unique information about their progenitors and jet launching mechanisms \citep{Nagakura2014JetBursts,Lazzati2017b, Urrutia2021,Gottlieb2021,Gottlieb2022,2023SciA....9I1405O}. 

A breakthrough came with the discovery of GRB\,170817A, the first binary neutron star merger detected via gravitational waves and electromagnetic radiation \citep{Abbott2017}. Unlike classical GRBs, this event was observed significantly off-axis, providing an unprecedented view of a relativistic jet from outside its core. The prompt gamma-ray signal was under-luminous compared to typical short GRBs \citep{Goldstein2017,ZhangBB2018NatCo}, while the afterglow exhibited a delayed rise \citep{troja2017,Hallinan2017} and a long-lived evolution \citep{Mooley2018, Troja2018_170817, Lazzati2018, Margutti2018,Lamb2019,troja2019,Troja2020,Balasubramanian2022,Ryan2024} 
- hallmarks of an off-axis jet. This provided the clearest observational evidence to date that GRB jets are not simple top-hat outflows, but instead possess an angular structure shaped by their interaction with the surrounding ejecta.

Besides the case of GRB\,170817A, structured or two-component jets -- consisting of a narrow ultra-relativistic core surrounded by a wider, slower outflow -- have been investigated to explain the complexity of some GRB afterglows \citep{Sheth2003,Peng2005,Racusin2008,2023SciA....9I1405O,Sato2025}, whereas the off-axis scenario was discussed to explain the peculiar behavior of some nearby and underluminous GRBs such 
as, for example, GRB 980425 \citep{Galama1998,Yamazaki2003}, GRB 031203 \citep{RamirezRuiz2005}, and GRB 171205A \citep{Delia2018}. 
However, the nature of these nearby sub-energetic bursts remains debated. 
In particular, it is still unclear whether they represent classical GRBs viewed off-axis or instead arise from a distinct population powered by a different central engine and involving mildly relativistic quasi-isotropic outflows \citep{Soderberg2006,Liang2007,Guetta2007,Virgili2008,Xu2023}.

\begin{figure*}
    \centering
    \includegraphics[width=1\linewidth]{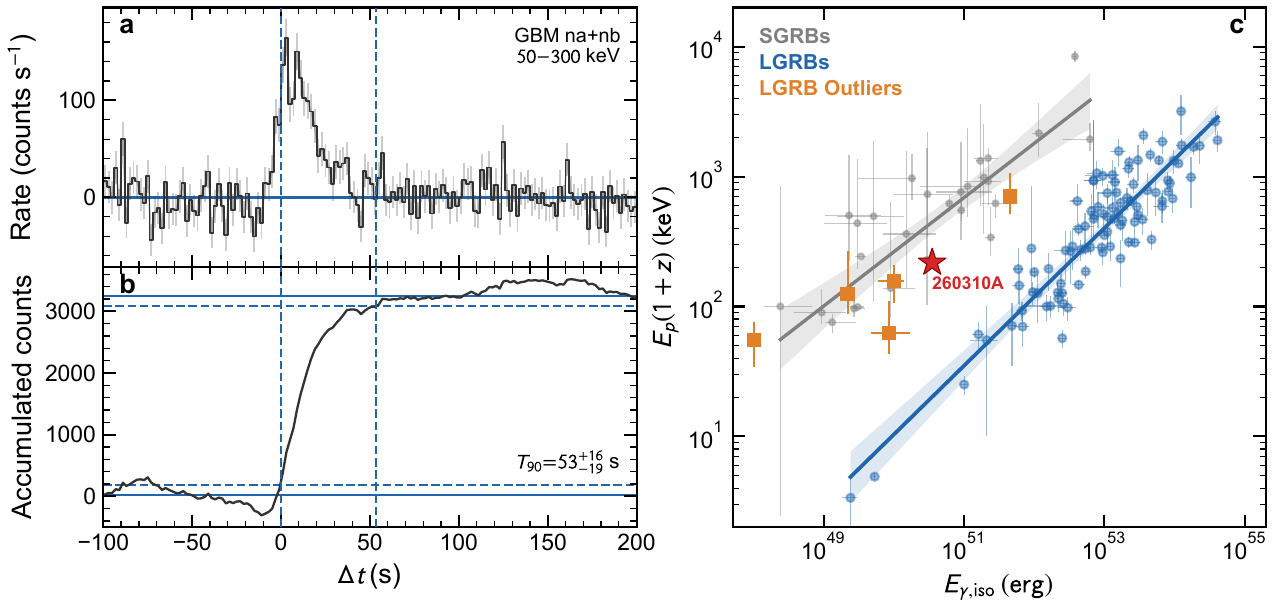}
    \caption{Prompt emission properties of GRB\,260310A. \textbf{a}.  \textit{Fermi}/GBM count-rate light curves from NaI detectors na and nb in the $50-300$ keV energy band. The black lines and gray lines represent the net counts and errors with a bin size of 2 s. \textbf{b}. Accumulated counts. The blue horizontal solid (dashed) lines are drawn at 0\%--100\% (5\%--95\%) of the total accumulated counts. The vertical lines in \textbf{a} and \textbf{b} represent the $T_{90}$ interval. \textbf{c}. Amati-relation diagram\citep{Amati2002,ZhangBB2018NatCo,ZhangBB2018NatAs}. Type I (short) and Type II (long) GRBs are shown as gray and blue circles, and the corresponding solid lines and dashed regions are the best-fit models and 95\% c.l., respectively. The red star marks the position of GRB\,260310A. Other outliers (GRBs 980425, 031203, 061021, 161219B, and 171205A) are marked as orange squares.  }
    \label{fig:gbmlc}
\end{figure*}

In this context, the recent GRB\,260310A is a particularly informative case.  
This burst, discovered by the Gamma-ray Burst Monitor \citep[GBM;][]{Meegan}
onboard \textit{Fermi} \citep{GCN43951}, was successfully localized by wide-field optical monitors \citep{GCNGOTO, GCNDDOTI} thanks to its delayed bright afterglow peak. 
Subsequent observations at X-ray, optical and radio wavelengths single it out as one of the brightest GRB afterglows detected thus far \citep{Perley2026}. 
The burst was initially proposed as a possible merger-driven long burst \citep{Gehrels2006,Rastinejad2022,Yang2022,Troja2022,Yang2024} due to its deviation from the canonical Amati relation \citep{Amati2002,GCN43981} and large galactocentric offset \citep{Bloom2002,Fong2022,OConnor2022}. However, despite these shared properties, GRB\,260310A has ultimately been identified as a standard long GRB followed by a broad-lined supernova at $z\approx$0.153 \citep{GCNdeUgartePostigo2026, GCN44137}.

In this paper, we present the results of our multi-wavelength follow-up observations, spanning $\sim 60$ days after the burst. 
Our dataset comprises extensive optical and near-infrared observations obtained with multiple facilities, radio observations with the Karl G. Jansky Very Large Array (VLA), together with X-ray monitoring by the Follow-up X-ray Telescope (FXT) aboard \textit{Einstein Probe} (EP), \textit{Chandra}, \textit{XMM-Newton} and Nuclear Spectroscopic Telescope Array (NuSTAR), tracing one of the brightest X-ray afterglows ever observed.

Throughout the manuscript we adopt a standard $\Lambda$CDM cosmology \citep{Planck2020} with $H_0$\,$=$\,$67.4$\,km\,s$^{-1}$\,Mpc$^{-1}$, $\Omega_\textrm{m}$\,$=$\,$0.315$, and $\Omega_\Lambda$\,$=$\,$0.685$. 
All the errors and upper limits are reported at the 1$\sigma$ and 3$\sigma$ confidence level, respectively.

\section{Observation and Data Reduction}

\subsection{Gamma-rays}
At UT 04:57:10 on March 10, 2026 (hereafter $T_0$), GRB\,260310A was triggered by the \textit{Fermi} Gamma-ray Burst Monitor (GBM) with an on-ground calculated location at RA = 213.6, Dec = 78.7 deg with a statistical uncertainty of 4.3 degrees \citep{GCN43951}. From the 12 sodium iodide (NaI) and 2 bismuth germanium oxide (BGO) detectors on board, we selected detectors na, nb, and b1, which provide the smallest viewing angles
relative to the GRB source direction. The time-tagged event data were retrieved and reduced with \textit{heapy}\footnote{\url{https://github.com/jyangch/heapy}}. The combined na and nb light curve of GRB\,260310A (Fig.~\ref{fig:gbmlc}, left panel) in the energy range of 50--300 keV exhibits a pulse with a fast-rise and exponential-decay profile, with a measured burst duration $T_{90} = 53_{-19}^{+16}$ s.
The minimum variability timescale \citep{Golkhou2014,Golkhou2015} is only loosely constrained to $\Delta t_{\min} <5$ s.

We conducted a time-integrated spectral fitting using na, nb and b1 data covering the $T_{90}$ time interval. The spectrum is well described by a cutoff power-law model with photon index of $-0.2\pm0.3$
and peak energy of $E_{\rm p}=157_{-13}^{+15}$ keV. 
The derived fluence is $5.2_{-0.3}^{+0.6}\times10^{-6}$ erg cm$^{-2}$  (10--1,000 keV), yielding a rest-frame isotropic-equivalent energy $E_{\gamma,{\rm iso}} = 3.5_{-0.2}^{+0.4}\times10^{50}$ erg. The derived peak flux and rest-frame peak luminosity in the time interval of $[T_0,~T_0+2 {\rm~s}]$ are $2.3_{-0.4}^{+0.6}\times10^{-7}$ erg cm$^{-2}$ s$^{-1}$ and $L_{\gamma,{\rm pk}}=1.5_{-0.3}^{+0.4}\times10^{49}$ erg s$^{-1}$.
The location of GRB\,260310A in the Amati diagram \citep{Amati2002} is shown in Fig.~\ref{fig:gbmlc}c.

\subsection{X-rays}\label{sec:xray}
\subsubsection{Einstein Probe}\label{sec:ep}
The position of GRB\,260310A was not covered by the Wide-field X-ray Telescope (WXT) onboard EP within a time window of $T_0\pm1$ hours around the GBM trigger. 

The Follow-up X-ray Telescope (FXT) onboard EP began follow-up observations of the field at UT 2026-03-12 17:50:08 ($\Delta t = 2.5$ d). An uncatalogued X-ray source was detected at R.A. = 219.3152 deg, Dec = 71.8417 deg (J2000) with an uncertainty of 10\arcsec~in radius (90\% confidence level; \citealt{GCN43994}). We further observed it through ToO requests (PIs: Yang, Jayaraman, Xu, Ghirlanda) for 19 epochs that spanned from March 14 to April 25, 2026.
All observations were conducted with a thin filter in full frame mode. The FXT data were processed using the FXT Data Analysis Software (\texttt{FXTDAS} v1.30), using the latest FXT calibration database (CALDB v1.30). Source photons were extracted from a circle with a radius of 60\arcsec~while background photons were extracted from an annulus with radii of 200--400\arcsec.

\begin{figure*}
    \includegraphics[width=1\linewidth]{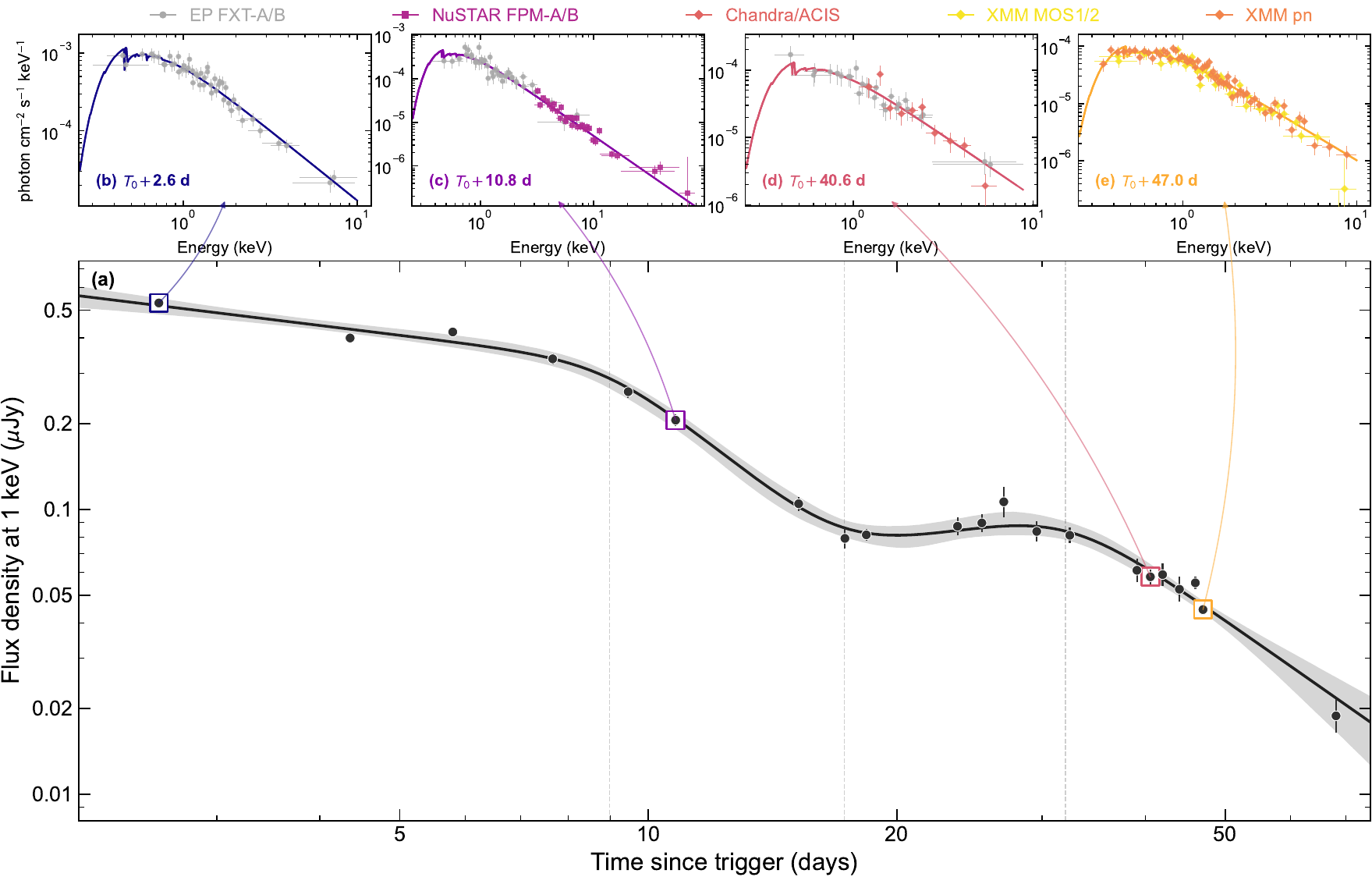}  
    \caption{X-ray observations of GRB\,260310A. 
    \textbf{a.} 1 keV flux density light curve fitted with a four-segment smoothly broken power-law model (black solid line) and the associated 95\% confidence region (gray shaded area). Vertical dashed lines indicate the break times.
    Detailed spectral views of four representative epochs:
    (\textbf{b}) $T_0 + 2.6$ d (the initial EP/FXT epoch), 
    (\textbf{c}) $T_0 + 10.8$ d (the first NuSTAR epoch with quasi-simultaneous FXT observations), 
    (\textbf{d}) $T_0 + 40.9$ d (the \textit{Chandra} epoch with quasi-simultaneous FXT observations), 
    and (\textbf{e}) $T_0 + 47$ d (the \textit{XMM-Newton} epoch). 
    The first six epochs (Interval 1) and the remaining epochs are joint-fitted using a tied intrinsic absorption and a shared photon index within each respective interval. Solid lines in the spectral panels illustrate the best-fit model for each epoch. Spectral fitting was performed on data grouped to $\ge 1$ count per bin using $C$-statistic and data are rebinned for display purposes.
    }
    \label{fig:x}
\end{figure*}

The resulting X-ray light curve is shown in Fig.~\ref{fig:x}. 
We describe the afterglow temporal evolution with a smoothly broken power-law composed of $n=4$ segments, defined as:
\begin{equation}
F(t) = F_0 t^{-\alpha_1} \prod_{k=1}^{n} \left[1 + \left(\frac{t}{t_{b,k}}\right)^{s}\right]^{-\Delta\alpha_k / s}
\label{eq}
\end{equation}
where $\alpha_k$ are the slopes,  $\Delta\alpha_k = \alpha_{k+1} - \alpha_k$, $t_{b,k}$ are the three temporal break times, and $s$ is a smoothness parameter that controls the sharpness of the transitions. Log-uniform priors were adopted for the flux normalization parameter $F_0$ and temporal breaks $t_{b,k} \in [1,70]$~d with $t_{b,1} < t_{b,2} < t_{b,3}$ enforced. Uniform priors are used for decay indices $\alpha_{k} \in [-5,5]$ and smoothness parameter $s \in [1,50]$.
Parameter estimation was performed with the nested-sampling algorithm via the Python package \texttt{PyMultiNest} \citep[used hereafter unless otherwise specified]{Buchner2014}.
The light curve (Fig.~\ref{fig:x}) exhibits a shallow decay with $\alpha_{X,1}=0.34\pm0.06$, followed by a steep decline with $\alpha_{X,2}=2.5^{+1.4}_{-0.6}$ after $t_{b,1}= T_0+9.1_{-0.8}^{+1.2}$~d. 
Between $t_{b,2}= T_0+18\pm2$~d and $t_{b,3}= T_0+30\pm3$~d, a bump is visible with a slope of $\alpha_{X,3}=-1.2^{+0.9}_{-2.0}$, which subsequently fades with $\alpha_{X,4}=2.1^{+0.4}_{-0.3}$.

\begin{figure*}
    \centering
    \includegraphics[width=0.45\linewidth]{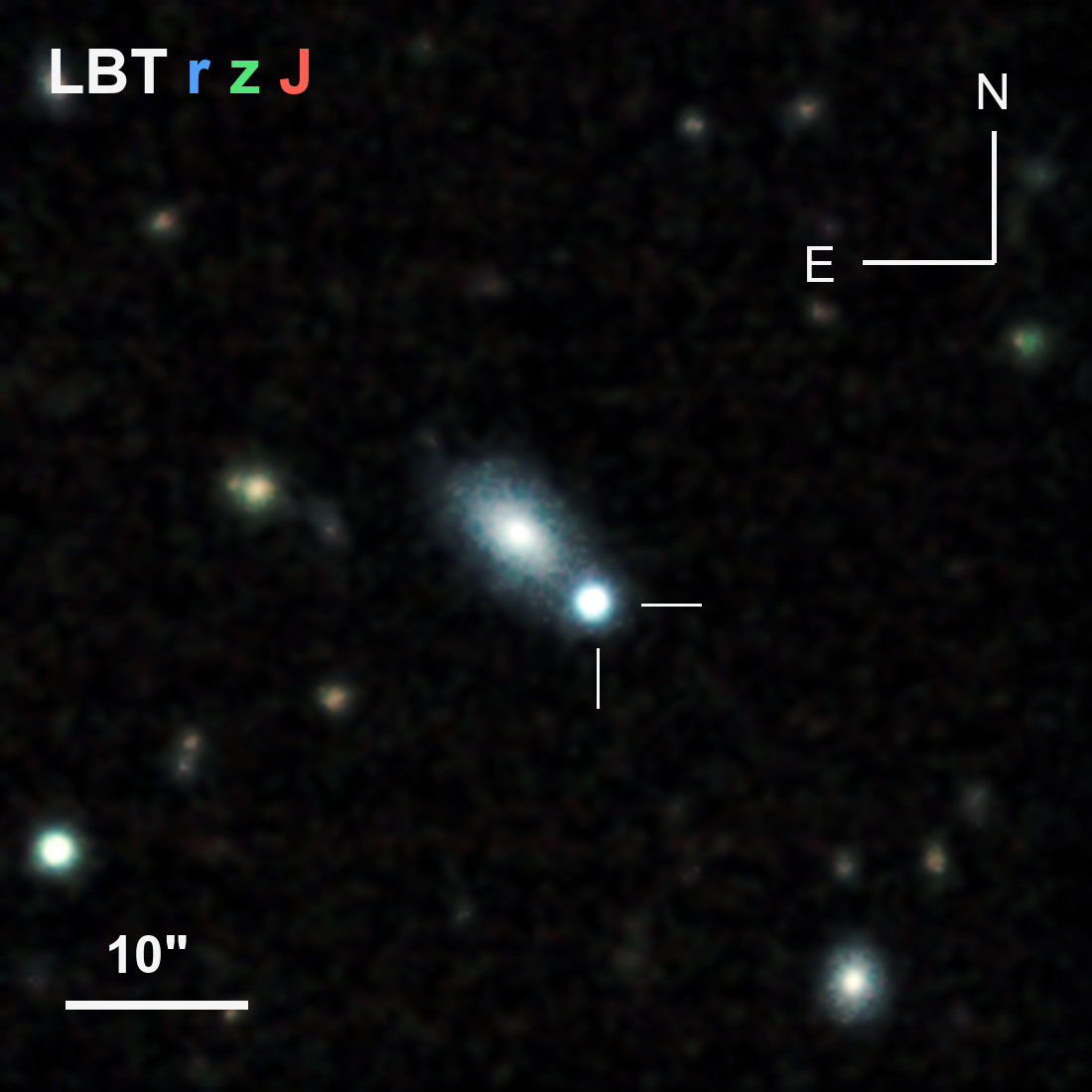}
    \hspace{0.2cm}
    \includegraphics[width=0.45\linewidth]{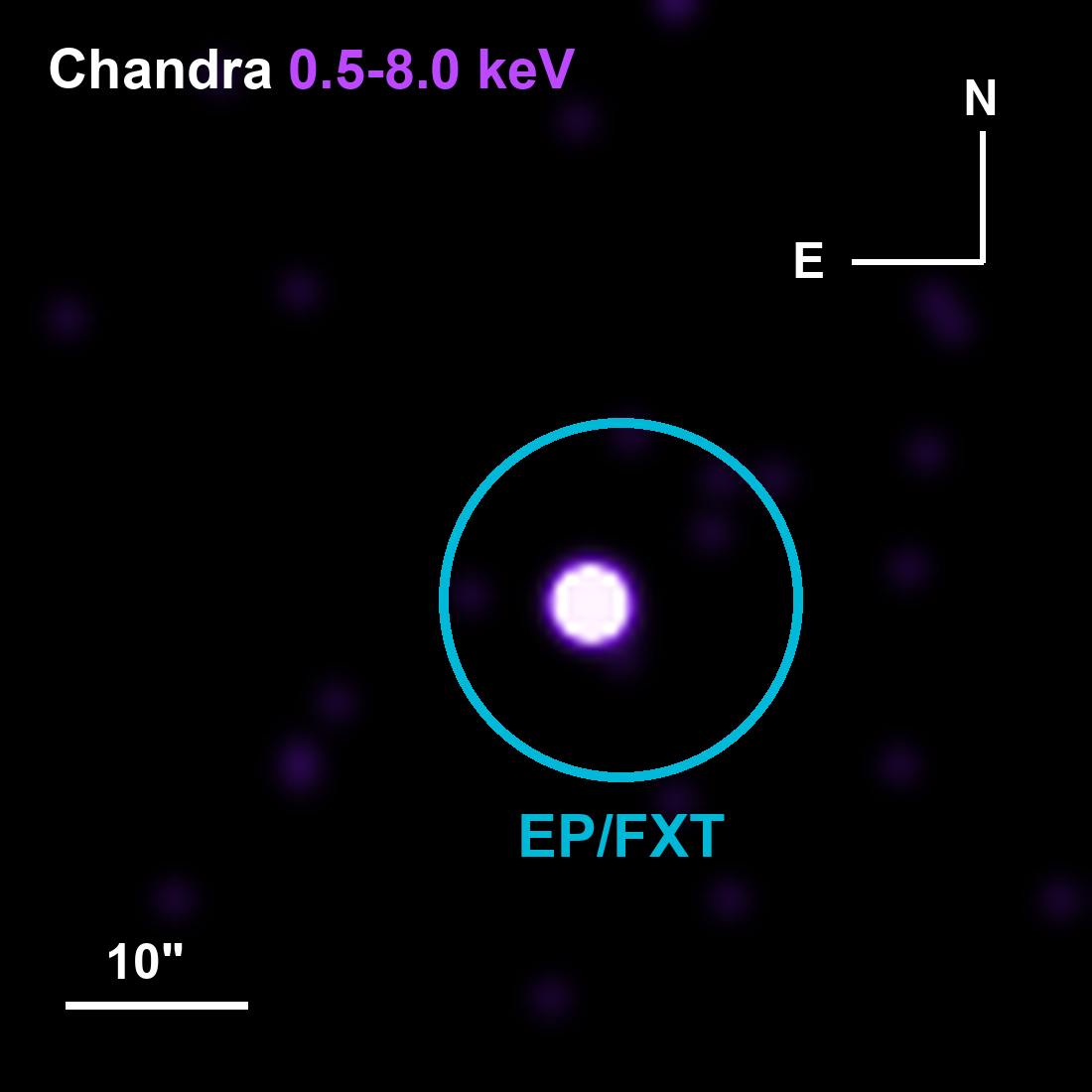}
    \caption{\textit{Left panel:}
False-color \added{Large Binocular Telescope (LBT)} image of the field of GRB\,260310A, constructed from the $r$, $z$, and $J$ bands, mapped to blue, green, and red, respectively. The transient position is marked by white ticks. 
\textit{Right panel:} 0.5-8.0 keV \textit{Chandra} image of the same field, smoothed with a Gaussian kernel of $\sigma$ = 2 pixels. The cyan circle marks the EP/FXT localization region. In both panels, north is up and east is to the left, and the scale bar is 10 arcsec.}
    \label{fig:image}
\end{figure*}

\subsubsection{Chandra X-ray Observatory}\label{sec:cxo}

To elucidate the origin of the late-time rebrightening visible in the EP lightcurve (Fig.~\ref{fig:x}), we initiated a Director's Discretionary Time (DDT) Target of Opportunity (ToO) observation with the \textit{Chandra} X-ray Observatory. Thanks to its superb angular resolution, only \textit{Chandra} can assess whether the X-ray measurements are contaminated by a nearby source (Fig.~\ref{fig:image}), unresolved within the EP/FXT's broader point spread function (PSF). 

The observation  (ObsID: 32306; PI: Yang) was performed at $\approx T_0 + 41$ d with the Advanced CCD Imaging Spectrometer (ACIS), with the target positioned on the back-illuminated S3 chip.
The data reduction was carried out using the Chandra Interactive Analysis of Observations software \citep[\texttt{CIAO v4.18};][]{CIAO} and the corresponding calibration database (\texttt{CALDB v4.12.3}). Level 2 event files were reprocessed using the \texttt{chandra\_repro} pipeline and cleaned from periods of elevated background. 
After standard filtering, the total cleaned exposure time amounts to $\sim$5 ks.
The absolute astrometry was corrected using \texttt{fine\_astro,} cross-matched against the Gaia DR3 catalog \citep{2023A&A...674A...1G}.

\textit{Chandra} imaging identifies the GRB afterglow as the dominant contributor to the total flux, and two other weaker sources detected within a 1\arcmin~circular region \citep{GCN44375}. This confirms that the observed X-ray rebrightening is an intrinsic feature of the GRB afterglow.

An additional \textit{Chandra} observation (ObsID: 32337; PI: Jayaraman) was carried out at $T_0 + 68$ d for an exposure time of $\sim$10 ks. The data were processed following the same reduction procedure described above. Spectra are extracted via \texttt{srcflux} using a 1\arcsec~circular region for the source, and a concentric 15--30\arcsec~annulus for the background.

\subsubsection{XMM-Newton}
After establishing that the late-time X-ray rebrightening is related to the GRB, we initiated a DDT ToO observation with {\itshape XMM-Newton}  (ObsID: 0974792201; PI: Yang) to constrain its late-time decay and spectral shape. 
The target was observed on April 26, 2026 ($T_0+47$~d)
using the three detectors (pn, MOS1 and MOS2) on the EPIC in full window mode, with the thin optical-blocking filter. {\itshape XMM-Newton} data was reduced by the \texttt{Science Analysis System (SAS)} v22.1 with the most recent calibration files. 
The exposure times, excluding high-rate flaring particle background, are 16~ks and 20~ks for the pn and MOS detectors, respectively. 
Source spectra were extracted using a 15\arcsec~circular region centered at the GRB coordinates, and the background was obtained using a 60--100\arcsec~annulus for both MOS detectors and a nearby source-free 40\arcsec~circle for the pn detector.

\subsubsection{Nuclear Spectroscopic Telescope Array}\label{sec:nustar}

NuSTAR \citep{Harrison2013} observed GRB\,260310A at two epochs ($T_0+10.7$~d and $T_0+32.4$~d) through a DDT ToO request (ObsIDs: 91202417002, 91202417004; PI: Waratkar). 
The preliminary results of these observations \citep{2026GCN44063, 2026GCN44278} report the detection of an X-ray counterpart in both epochs.

The NuSTAR data were reduced using the standard tasks in the NuSTAR Data Analysis Software pipeline (\texttt{NuSTARDAS}), distributed as part of \texttt{HEASoft} v6.34.
We extract the source and background spectra using \texttt{nuproducts} from a 50\arcsec~circular region and a 120$-$200\arcsec~annulus centered on the source, respectively, for both FPMA and FPMB.

\subsubsection{X-ray Spectral Analysis}
Motivated by the changes in temporal behavior (\S\ref{sec:ep}), we divided the X-ray observations into two temporal intervals and analyzed them independently. Interval 1 covers the period from $T_0+2.6$ d to $T_0+11$ d and includes the first six FXT epochs together with the first NuSTAR epoch, while interval 2 spans from $T_0+15$ d to $T_0+47$ d and comprises the remaining FXT epochs, the \textit{Chandra} and \textit{XMM-Newton} observations, as well as the second NuSTAR epoch.

The X-ray spectra, obtained with FXT (0.3--10 keV), \textit{Chandra} (0.5--8 keV), NuSTAR (3--79 keV), and \textit{XMM-Newton} (0.3--10 keV) were fitted using the \texttt{XSPEC} v12.15 \citep{XSPEC} with an absorbed power-law model considering a Galactic hydrogen column density $N_{\rm H} = 2.59 \times 10^{20}$~cm$^{-2}$ \citep{Willingale2013} and the intrinsic absorption $N_{\rm H, z}$ from the GRB site \citep[\texttt{tbabs*ztbabs*powerlaw};][]{Wilms2000ApJ}. 
The spectra are well fit with a common 
$N_{\rm H,z} = (1.6 \pm0.2)\times 10^{21}$~cm$^{-2}$
and consistent photon indices $\Gamma = 1.80\pm0.04$ (interval 1) 
and $\Gamma = 1.82\pm0.05$ (interval 2)\footnote{
Uncertainties are quoted for the joint 1$\sigma$ confidence regions for the three parameters \citep{Lampton1976}, accounting for the inherent correlations among them.
}, 
indicating no evidence for spectral variation
within the uncertainties. 

Our results differ from the preliminary report of \citet{GCN43994} with $\Gamma \approx 1.55$. This discrepancy can be explained by the different absorption model adopted in \citet{GCN43994}, which reports only a fixed Galactic hydrogen column density, whereas our fit includes both Galactic and intrinsic absorption components.

The unabsorbed flux in the 0.3--10 keV band and the flux density at 1 keV (with the 3--79 keV fluxes and 10 keV flux density additionally reported for NuSTAR), derived from these best-fit models, are listed in Tab.~\ref{tab:obslog} and shown in Fig.~\ref{fig:x}.

\begin{figure*}
    \centering
    \includegraphics[width=\linewidth]{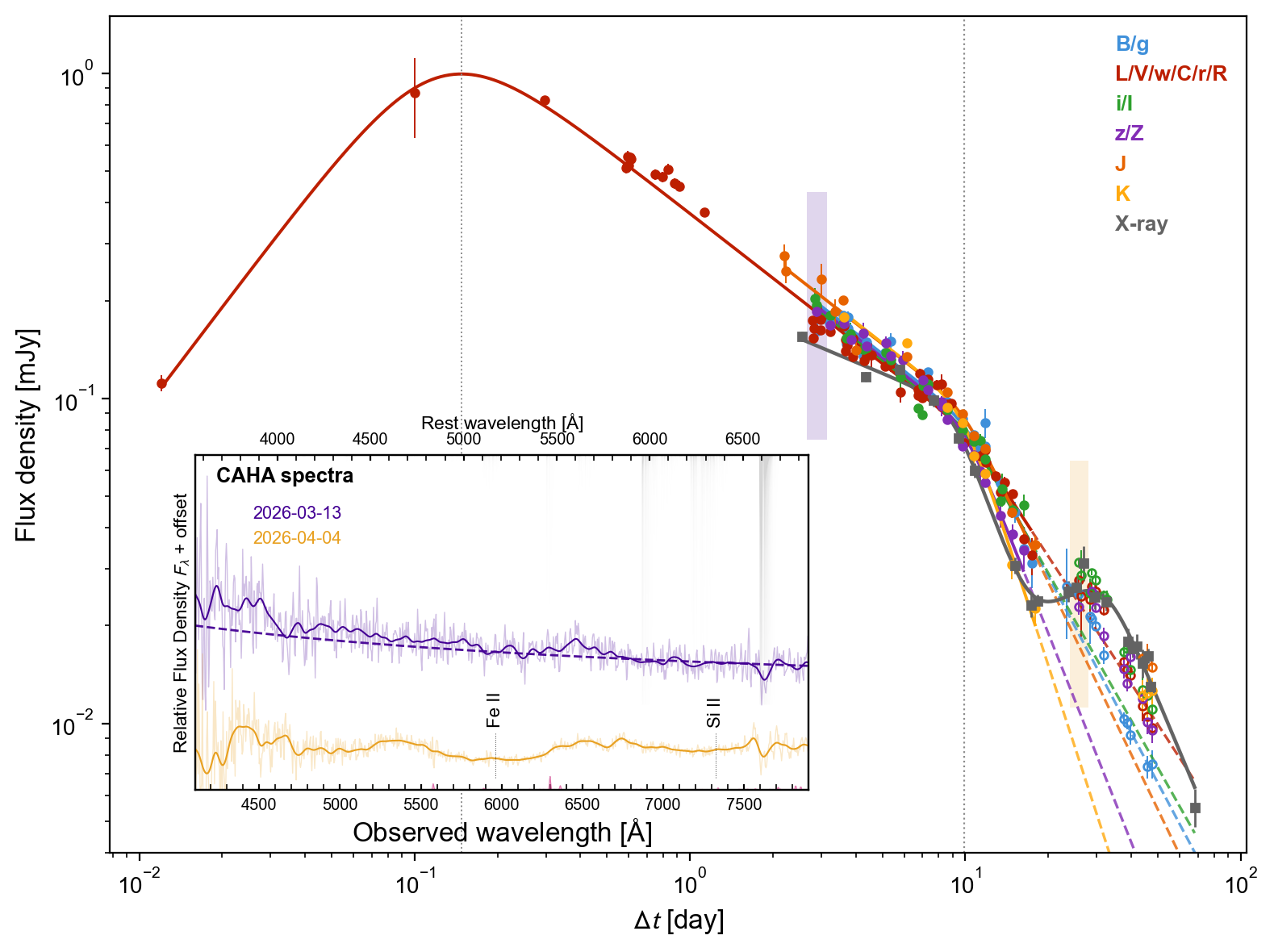}
    \caption{Multi-color optical and nIR lightcurves of GRB\,260310A, compared to the X-ray lightcurve (gray squares). \added{Data were rescaled to the $R$ band assuming $F_\nu \propto \nu^{-\beta}$ with $\beta = 0.91$. 
    The absolute X-ray flux is shown in Fig.~\ref{fig:x}a.} 
    The red solid line shows the best fit temporal model, describing the early optical data.
    Dashed lines show the late-time decay in different colors. 
    Vertical bars mark the two epochs of optical spectroscopy. 
    \textit{Inset:} CAHA spectra observed on March 13 and April 4, 2026, corrected for Galactic extinction. To guide the eye, the dashed line shows a power-law spectrum with slope 0.91. The observed spectra are plotted as semi-transparent lines, with smoothed solid lines for plotting purposes. Smoothing is performed using a Gaussian kernel with a standard deviation of 5 pixels.   
    Wavelengths affected by telluric absorption and sky emission are marked by gray and pink areas, respectively \citep{Lord1992}.
    }
    \label{fig:opt}
\end{figure*}

\subsection{Optical and Near-Infrared Imaging}\label{sec:ONII}
We carried out imaging observations of the field of GRB\,260310A across a broad range of optical and near-infrared bands using multiple facilities: 
the Burst Observer and Optical Transient Exploring System \citep[BOOTES;][]{2023NatAs...7.1136C,HuYD2023}, the Observatorio de Sierra Nevada (OSN), the Large Binocular Telescope (LBT), the Ond\v{r}ejov observatory, and the Optical Monitor (OM) onboard \textit{XMM-Newton}.

The transient is located in the outskirts of a bright galaxy (Fig.~\ref{fig:image}),
whose diffuse light partially affects the target's photometry. 
Archival $grz$-band images from the DESI Legacy Imaging Surveys \citep{Dey2019} and $i$-band images from the Pan-STARRS Surveys \citep{Chambers2016,Magnier2020}
served as reference templates. 
Unless otherwise specified, aperture photometry for the optical and NIR imaging was calibrated against the ATLAS All-Sky Stellar Reference Catalog \citep[Refcat2;][]{2018ApJ...867..105T} and the Two Micron All Sky Survey \citep[2MASS;][]{2mass} catalog, respectively. 
Further details are provided in Appendix~\ref{sec:optical_details}. The photometric results for the complete dataset, including supplementary optical/NIR data obtained from public archives, are presented in Table~\ref{tab:obslog}. Additional optical/NIR data from \citet{OConnor2026} were also incorporated into the analysis. Galactic extinction corrections were applied to the entire dataset prior to model fitting using the \texttt{G23} extinction curve model \citep[][see also \citealt{Gordon2009}, \citealt{Fitzpatrick2019}, \citealt{Gordon2021}, and \citealt{Decleir2022}]{Gordon2023}, adopting $R_V=3.1$ and $A_V=0.058$ from the \textit{Gaia} TGE map \citep{Delchambre2023}.

After excluding the rebrightening at $>20$ d, we fitted the optical/nIR light curves with a smoothly broken power-law model (Eq.~\ref{eq}) 
composed of $n=3$ segments. 
The first two temporal slopes were tied across all bands, while the third slope was allowed to vary independently. A systematic uncertainty of $5\%$ was added in quadrature to the photometric errors. 
We fixed the smoothness parameter to a relatively soft value of $s=3$ for the first break, which provides a better representation of the broad optical peak, while a sharp $s=10$ was used for the latter break to better constrain the slopes.
The data and best-fit light curves, normalized to the $R$ band, are shown in Fig.~\ref{fig:opt}.
The optical/NIR emission initially rises with a temporal slope of $\alpha_1 \approx -1$ up to $\approx0.1$ d, followed by a shallower decay with $\alpha_2 = 0.67 \pm 0.01$. At $\sim$\,9 d, the light curves exhibit an achromatic break, also visible at X-ray energies. After this time, the flux decays at a faster rate with slope which varies among different bands, as also found by \citet{OConnor2026}. 
We derive $\alpha_3=1.7\pm0.2$ for $g/B$, $1.4\pm0.1$ for $L/V/w/C/r/R$, $1.5\pm0.2$ for $i/I$, $2.2\pm0.2$ for $z/Z$, $1.9\pm0.2$ for $J$, and $2.8\pm0.3$ for $K$ band. 
This chromatic behavior is likely due to the SN onset \citep{OConnor2026}.

\subsubsection{Dust Extinction}
Dust extinction alters the afterglow spectrum, introducing curvature relative to the intrinsic power-law shape.
To decouple the intrinsic optical/NIR spectral index from dust-induced curvature, we model the afterglow emission as $F_\nu(\nu,t) \propto \nu^{-\beta} t^{-\alpha}$ introducing an extinction component parametrized by the reddening $E(B-V)_z$.
We jointly fit the host-subtracted optical and nIR data (from this work, supplemented by \citealt{OConnor2026}) spanning $T_0 + 2$~d to $T_0 + 6$~d, during which no significant spectral or temporal breaks are observed.
All measurements were first corrected for Galactic extinction. 
The intrinsic dust extinction is modeled using the \texttt{G23} extinction curve with $R_V = 3.1$. Uniform priors are assigned to $\alpha$, $\beta$, and $E(B-V)_z$, and a log-uniform prior to the normalization. 
The fit constrains $\alpha = 0.65\pm0.05$ and $\beta_O = 0.91_{-0.09}^{+0.06}$ with reddening $E(B-V)_z = 0.02^{+0.03}_{-0.01}$, indicating a dust-poor local environment, consistent with the large galacto-centric offset of the transient.

When compared with the absorption column inferred from X-ray spectroscopy, the derived extinction implies a high gas-to-dust ratio, $N_{H,z}/A_V \approx2.5\times10^{22} {\rm\ cm}^{-2}\ {\rm mag}^{-1}$, an order of magnitude higher than in the Milky Way. 
This is consistent with the well-known trend of GRB sightlines showing large X-ray columns and modest optical reddening \citep{Stratta2004}.

\subsubsection{Optical Polarimetry}\label{sec:opol}

We conducted optical polarimetry with the DIPOL instrument installed at the 90 cm telescope (T90) in Observatorio de Sierra Nevada (Granada, Spain) in two epochs during the night of March 12 and March 13 (16$\times$300\,s each). 
Data were reduced using the Interactive Optical Photo-Polarimetric Python Pipeline \citep[IOP4;][]{iop4}. Resulting polarization was 2$\pm$2\% and 4$\pm$2\%, respectively, 
consistent with null polarization and in agreement with the  3$\sigma$ upper limit of $<$1.5\% reported at a similar epoch \citep{GCN43990}.

\subsection{Optical Spectroscopy}
Spectroscopic observations were carried out with the 2.2 m Calar Alto Faint Object Spectrograph (CAFOS) mounted on the 2.2 m telescope at Calar Alto Observatory (CAHA), Spain, on 2026 March 13 ($T_0+2.9$ d) and April 4 ($T_0+26$ d) under Program 26A-2.2-026 (PI: A. J. Castro-Tirado). Exposure times were $400$ s and $3\times900$ s for the two epochs, respectively. The observations employed a 2.9\arcsec\ slit together with the g200 grism, providing wavelength coverage over 4000--8000 \AA.
The first epoch was obtained under poor seeing and high-humidity conditions, while the second epoch was conducted under improved conditions, with a seeing of 1.2\arcsec.
The data were reduced following standard \texttt{IRAF} procedures, including bias, flat, wavelength, and flux calibrations.
The reduced spectra are shown in the inset of Fig.~\ref{fig:opt}.
In the first epoch, the spectrum displays 
a featureless power-law continuum consistent with afterglow-dominated emission. 
By the second epoch, the spectrum reveals the emergence of broad undulations. These spectral features are typical of broad-lined Type Ic supernovae (Ic-BL SNe), in which high ejecta expansion velocities broaden and blend atomic transitions \citep{Modjaz2016}.

Two broad bumps are visible around rest-frame 4600 \AA\ and 5800 \AA. 
The $\sim$4600 \AA\ rest-frame peak is dominated by the blended emission of Fe II multiplets, while the adjacent dips are formed by the strongly blueshifted, broad absorption components of Fe II $\lambda$5169 and Si II $\lambda$6355. 

Although the optical and X-ray light curves exhibit a rebrightening on comparable timescales (Fig.~\ref{fig:opt}), their spectral properties are markedly different and point to distinct emission components. Our spectral analysis of the X-ray rebrightening shows that it is well described by a non-thermal power-law spectrum (Fig.~\ref{fig:x}). In contrast, the optical rebrightening is spectroscopically classified as a Type Ic SN. This implies that the power-law component responsible for the X-ray rebrightening becomes subdominant in the optical band.

\subsection{Radio}
Observations with the Karl G. Jansky Very Large Array (VLA) in A-array were initiated on March 18th 2026 ($T_0+8.2$ d), under project 26A-516 (PI: Troja) in C-band with central frequencies of 6~GHz and corresponding bandwidths of 4~GHz. After the association with a SN became apparent, observations continued under Director's Discretionary Time (DDT; PI: D. Perley) and were publicly released. 
The flux, bandpass, and polarization angle calibrator was 3C286, the phase calibrators were J1531+7206 and J1357+7643, respectively. 

Data were retrieved from the National Radio Astronomy Observatory (NRAO) online archive and  flagged, calibrated, and imaged in Common Astronomy Software Applications \citep[CASA;][]{CASA2022} v6.6.1 using standard procedures. The imaging was performed with the \texttt{TCLEAN} task using a Briggs parameter of 0.5. The \texttt{IMSTAT} task was used to evaluate RMS noise in a region of the restored map away from bright side-sources.
A 2D Gaussian fit was performed inside the \texttt{viewer} task. The peak fluxes and the corresponding 1$\sigma$ uncertainties are reported in Table~\ref{tab:obslog}, and are consistent with those reported in the GCNs \citep{Perley2026, Schroeder2026}.

\subsubsection{Radio Polarimetry} 

Given the brightness of its radio counterpart, GRB260310A was an excellent target for polarization studies \citep[e.g.][]{Christy2026}. Unfortunately, the scheduling blocks discussed above relied on a polarized phase calibrator observed over a limited range of parallactic-angles. Such a setup does not allow to estimate the instrumental leakage terms (D-terms) necessary for a reliable calibration of the polarization vectors.

To recover a polarization model for the phase calibrator, we reobserved GRB\,260310A with VLA on May 5th  in Ku-band ($T_0 + 55$ d) and on May 17th in C- and X-band ($T_0 + 68$ d) adding an unpolarised calibrator (J1407+2827) to the schedule. 

We followed the standard polarization calibration steps: (1) setting the polarization model for the primary calibrator 3C286 \citep{Perley2017}; (2) correcting for cross-hand (RL, LR) delays to ensure phases are aligned using 3C286; (3) solving for instrumental leakage terms (D-terms) using the unpolarized calibrator J1407+2827; (4) solving for the R-L polarization phase difference using 3C286. The calibration solutions were then applied to the phase calibrator J1357+7643 and the GRB.

\begin{figure}
    \centering
    \includegraphics[width=1\linewidth]{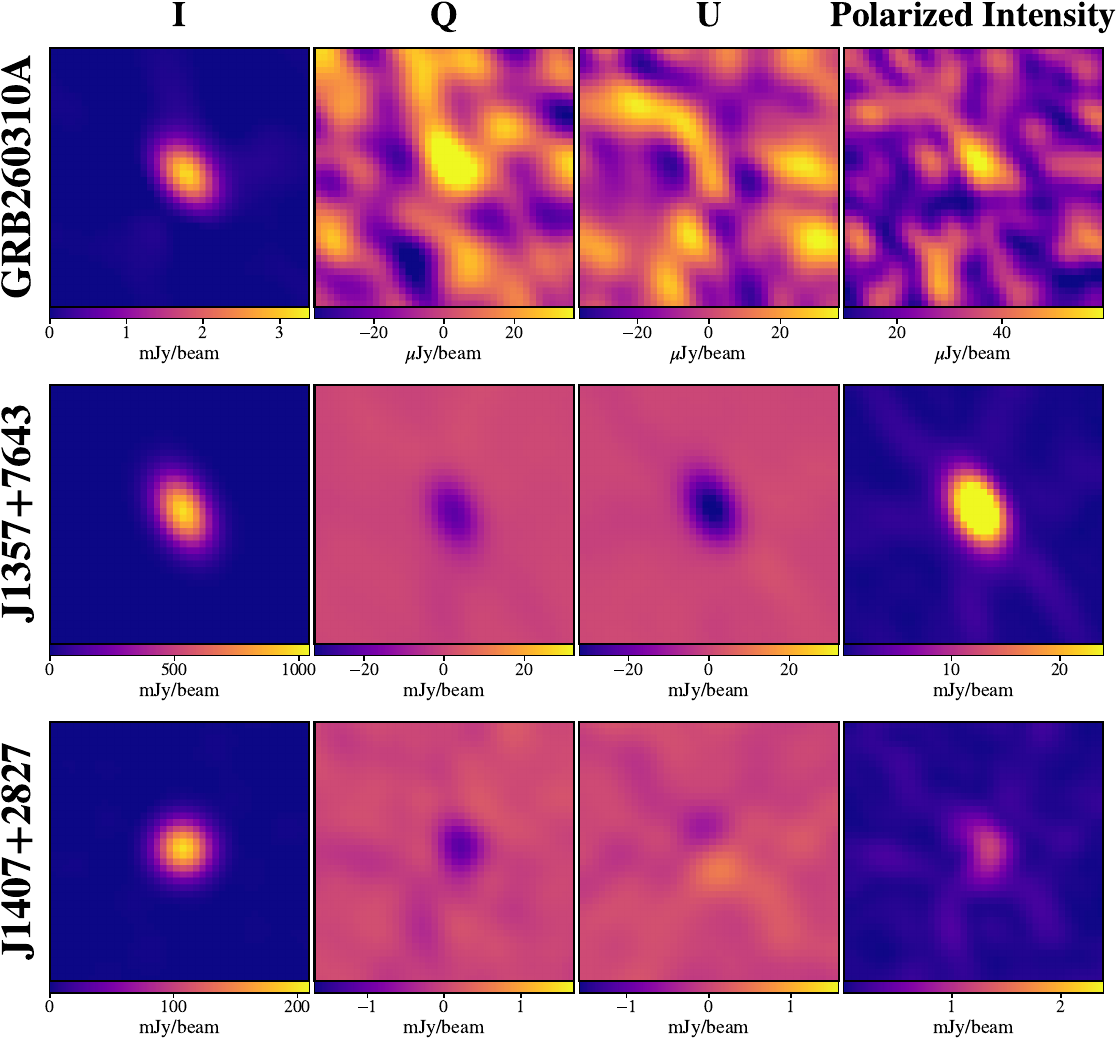}
    \caption{Stokes $IQU$ images and linear polarization map for the target GRB\,260310A (top), the phase calibrator J+1357 (middle) and the leakage term calibrator J+1407 (bottom).  The field was imaged at $T_0+$55 d in the Ku-band.}
    \label{fig:pol}
\end{figure}

The polarization-calibrated data of the primary calibrator,  phase calibrator and target were imaged using \texttt{TCLEAN} with all four Stokes parameters (I, Q, U, V) using a Briggs parameter of 0.5. 
The I-, Q- and U-Stokes Ku-band (15 GHz) images are shown in Fig.~\ref{fig:pol}.
From the Q- and U-Stokes maps the debiased linear polarization intensity was derived as $P = \sqrt{Q^2 + U^2 - \sigma_{\rm Q}^2}$ \citep{wardlekronberg1974}, where Q and U are measured at the position of the flux peak, and $\sigma^2$ is the image rms. 
The fractional polarization was then derived as $\Pi = P/I$, where I is the peak flux density of total intensity (I-Stokes). The polarization angle (PA) was calculated as \mbox{$\chi$ = 0.5 tan$^{-1}$(U/Q)}. 
The leakage terms calibrator showed a polarization fraction below 1\%  and the phase calibrator J1357+7643 was polarized  at a level of $\Pi = 4.1 \pm 0.2 \%$ with PA = $29.2 \pm 1.4 $ deg. For the target, we measure a polarization fraction $\Pi = 1.7 \pm 0.4 \%$ and derive an upper limit $|\rm{PA}| < 18$ deg. Uncertainties were computed by summing in quadrature a systematic term of 5\% of the I-, Q-, and U-Stokes peak flux values with the statistical term.  At lower frequencies we derive 3$\sigma$ upper limits $\Pi_{\rm 6 GHz} <$ 1.7\% and $\Pi_{\rm 10 GHz} <$ 2.5\%.

The polarization model of J1357+7643 was then applied to the entire dataset using the standard pipeline with the \texttt{Df+X} option. Unfortunately, the resulting calibration solutions yielded unrealistically large and highly scattered D-term values, indicating that the polarization calibration was unreliable. As a consequence, we were unable to derive meaningful polarization measurements from the earlier dataset.

\begin{figure}
    \centering
    \includegraphics[width=1\linewidth]{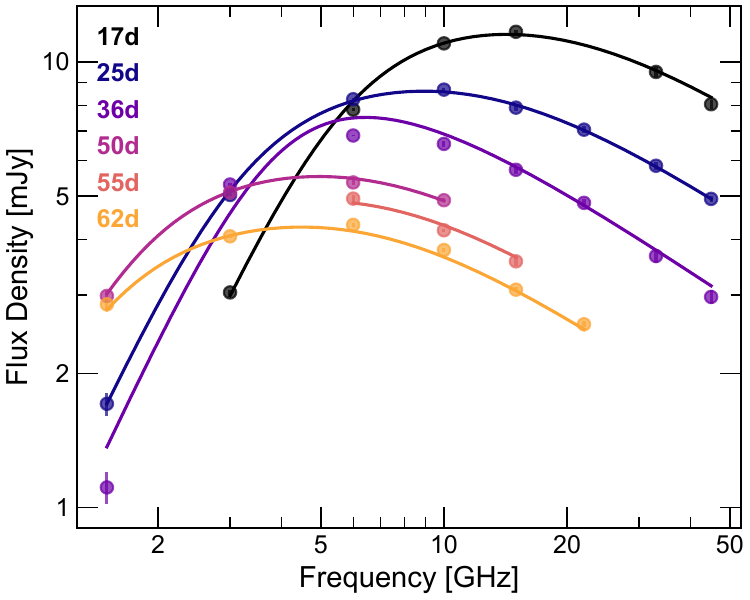}
    \caption{Radio spectral energy distributions (SEDs) of GRB\,260310A at five epochs ($T_0+$17, 25, 36, 50, 55 and 62~d, color-coded from dark to light). Solid curves are the best fit SEDs assuming a smoothness parameter $s=2$.}
    \label{fig:radioSED}
\end{figure}

\subsubsection{Radio Spectral Analysis}
\label{sec:radiosed}

Dense sampling of the radio spectrum starts at $T_0 + 17$~d, right after the rapid decay of the optical and X-ray afterglows. In this phase, the optical emission begins to be dominated by the SN light, whereas the X-rays show an afterglow rebrightening. 

As shown in Fig.~\ref{fig:radioSED}, at low frequencies ($\lesssim$4 GHz), the radio flux shows a steep rise, characteristic of an optically thick spectrum. At higher frequencies, the spectral shape is instead consistent with optically thin emission. 
We model the radio SED using a smoothly broken power-law based on a synchrotron spectrum. The function consists of three power-law segments with slopes of $-2$, $-1/3$, and $(p-1)/2$, separated by two spectral breaks corresponding to the self-absorption frequency $\nu_a$ and the characteristic synchrotron frequency $\nu_m$, with $\nu_m > \nu_a$. A smoothness parameter $s$ controls the sharpness of the transitions between the different segments.

We simultaneously fit the data from six epochs ($T_0+$17, 25, 36, 50, 55 and 62~d), allowing $\nu_a$, $\nu_m$, and the normalization to vary between epochs. We adopted log-uniform priors for $\nu_a$ and $\nu_m$, and a uniform prior on the electron power-law index $p$ over the range $[2,3]$. The parameter $p$ was assumed to be common to all epochs, while the smoothness parameter was fixed. 
For the two adjacent epochs of 50 and 55~d, we additionally imposed a common value of $\nu_a$ and $\nu_m$ to better constrain the fit using their complementary spectral coverage.
The likelihood incorporates a 5\% systematic calibration error added in quadrature to the statistical flux uncertainties.

The radio SED  and best fit models for $s=2$ are shown in Fig.~\ref{fig:radioSED}. From the fit, we derive $2\lesssim p \lesssim 2.3$, depending on the adopted smoothness parameter. 
The peak flux decays with time as $\propto t^{-0.7}$, whereas the characteristic frequencies decrease as $\nu_m \sim \nu_{m,0} \, t^{-1.42}$ and $\nu_a \sim \nu_{a,0} \, t^{-0.54}$, where $\nu_{m,0}\approx$\,17~GHz and $\nu_{a,0}\approx$\,5~GHz at 17 d. 
At $T_0 + 50$~d, the derived values of $\nu_m \approx$\,6~GHz and $\nu_a \approx$\,2~GHz deviate from this trend.

\begin{figure*}
  \centering
    \includegraphics[width=0.93\linewidth]{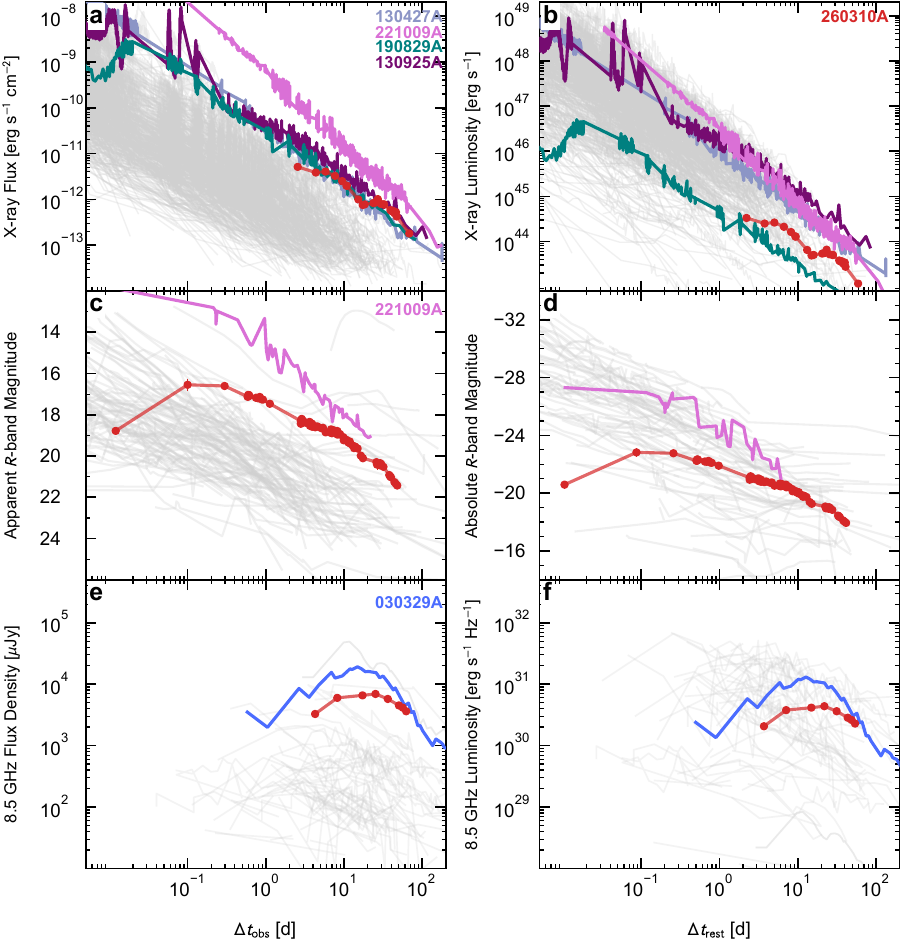}
    \caption{Comparison of GRB\,260310A with the broader population of GRB afterglows with known redshifts. The panels display the observer-frame X-ray flux (0.3-10 keV; top left), rest-frame X-ray luminosity (0.3-10 keV; top right), optical apparent magnitude (R-band; middle left), optical absolute magnitude (R-band; middle right), radio observed flux density (8.5\,GHz; bottom left) and radio rest-frame luminosity (8.5GHz; bottom right). 
    }
    \label{fig:comp}  
\end{figure*}

\begin{figure*}
    \centering
    \includegraphics[width=\linewidth]{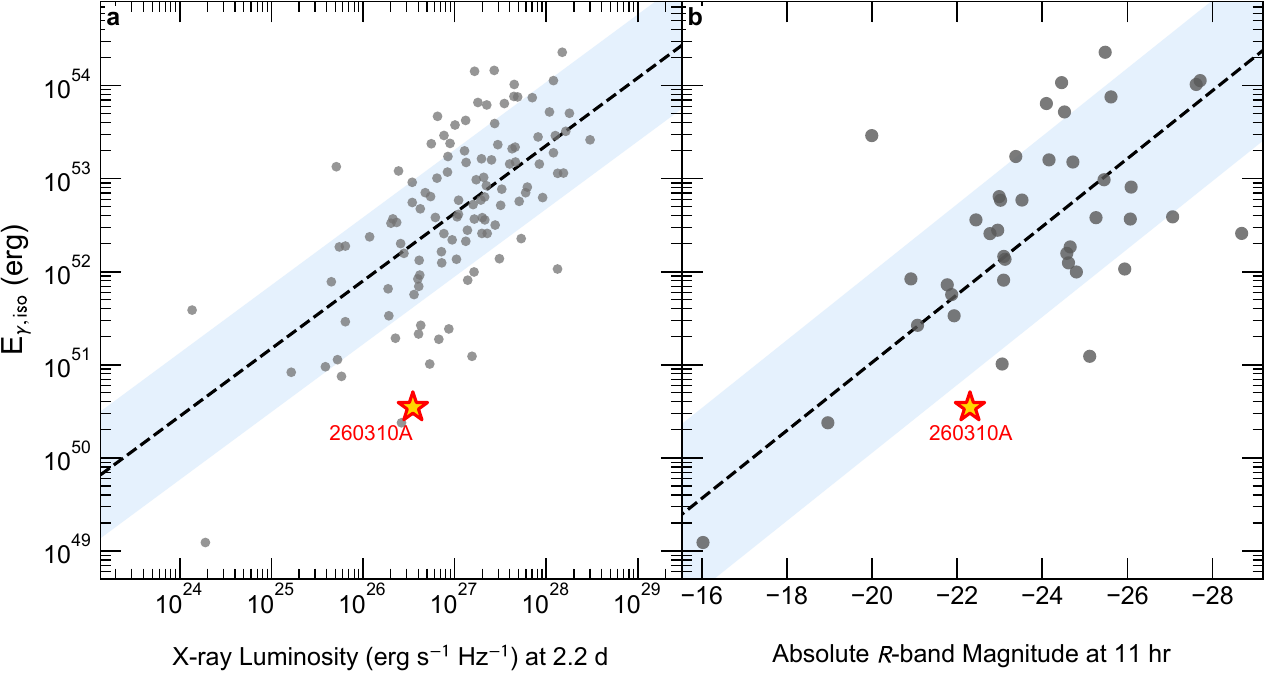}
    \caption{Afterglow rest-frame X-ray luminosity at rest-frame 2.2 d (\textbf{a}) and $R$-band absolute magnitude $M_R$ at rest-frame 11 hr (\textbf{b}) versus 10–1000 keV prompt emission isotropic-equivalent energy for a sample of GRBs detected by \textit{Fermi}/GBM. The red-edged star highlights GRB\,260310A. Black dashed lines and shaded area represent the best-fit log-linear correlations and the corresponding 1$\sigma$ bands.
    }
    \label{fig:ER}
\end{figure*}

\section{Discussion}

\subsection{GRB\,260310A in context}\label{sec:context}

In Fig.~\ref{fig:comp} (left), we compare the X-ray (top), optical (middle) and radio (bottom) afterglows of GRB\,260310A with those of GRBs with known redshifts.  Our comparison sample includes 485 X-ray lightcurves from the \textit{Swift} XRT Database \citep{Evans2009}, 467 optical lightcurves from the repository of \citet{Dainotti2024}\footnote{\url{https://github.com/SLAC-Gamma-Rays/grbLC}}, and 64 events from the radio sample of \citet{Dichiara2022} and \citet{Gulati2026}\footnote{\url{https://github.com/ashnagulati/Transient_Comparison_Plots}}. 

To construct the X-ray light curves, we retrieve the count-rate data and the corresponding spectral information (unabsorbed counts-to-flux ratio and photon index) from the \textit{Swift} online repository\footnote{\url{https://www.swift.ac.uk/xrt_products}}. The unabsorbed flux light curves are derived by applying the corresponding time-averaged counts-to-flux conversion factors, then transformed into rest-frame luminosity light curves by accounting for cosmological time dilation and applying the appropriate k-correction \citep[e.g.][]{Bloom2001,Mangano2006}. 

To derive the optical light curves, we use the compilation of 535 GRBs with optical afterglow data \citep[][corrected for Galactic extinction]{Dainotti2024}, and filter for valid photometric measurements near the $R$ band. This selection yields a sub-sample of 467 light curves in the observer's frame, as displayed in Fig.~\ref{fig:comp}c. 
To derive the rest-frame $R$-band light curves (Fig.~\ref{fig:comp}d), the observed magnitudes from different filters are shifted to the rest-frame $R$ band and converted into absolute magnitudes. This step accounts for distance modulus correction, k-correction \citep[e.g.][which incorporates a redshift-dependent term and a spectral-index-dependent color-term]{Kann2010} and for cosmological time dilation. In cases where the spectral index is not tabulated, the color correction is omitted only for observed wavelengths sufficiently close to the rest-frame $R$ band. This yields a sample of 368 standardized light curves. 
A similar approach is adopted to build the rest-frame radio lightcurves \citep[e.g.][]{Chandra2012,Dichiara2022}.

In the observer's frame, GRB\,260310A stands out at all wavelengths as one of the brightest afterglows ever detected: 
its X-ray flux lies in the top 1\% of the distribution,  its optical emission peaks at $\approx$16.5 AB mag at $\approx$0.2 days and then surpasses in brightness 99\% of the sample, 
whereas its radio counterpart also tops the distribution of observed fluxes (brighter than 95\% of the sample). 
However, this extreme apparent brightness is largely driven by its proximity. At a redshift of $z\approx0.153$, GRB\,260310A lies within the closest $\sim3\%$ of the GRB population, which significantly boosts its observed flux across all bands.

In the rest-frame (Fig.~\ref{fig:comp}, right column), GRB\,260310A falls instead well within the broad distribution of cosmological GRBs. Its X-ray, optical, and radio luminosities cluster around the median values of the population, within a factor of a few of the typical afterglow luminosities, indicating no extreme properties once distance effects are removed.
Both its time-averaged and peak radio luminosities are consistent with the median of the population.
Its peak optical magnitude is fainter than roughly 80\% of the sample, while its X-ray luminosity lies in the lower 15\% of the distribution. It is only at later times, following its X-ray rebrightening, that its X-ray luminosity approaches extreme events such as GRB 221009A \citep{Williams2023,2023SciA....9I1405O} and GRB 130427A \citep{Maselli2014}.

Nevertheless, its apparent brightness is not its only peculiar feature. 
GRB afterglow brightness shows a broad correlation with the prompt gamma-ray energy output \citep{Berger2007,Davanzo2012}, reflecting the common origin in the relativistic flow.  GRB\,260310A instead exhibits a relatively modest gamma-ray fluence $5.2_{-0.3}^{+0.6}\times10^{-6}$ erg cm$^{-2}$, placing it at the 66\% of GBM GRBs,
yet displays with one of the brightest optical, radio and X-ray afterglows. 
As shown in Fig.~\ref{fig:ER}, this clearly separates it from the bulk of the GRB population, and is reminiscent of the behavior of GRB 150101B, a short GRB likely seen off-axis \citep{Troja2018b}. 
In this Figure we consider a sample of 112  GRBs jointly detected by \textit{Swift} and \textit{Fermi}, which are characterized by a broadband prompt emission spectrum to infer $E_{\gamma,\rm iso}$ along with afterglow measurements to probe the blastwave kinetic energy. 
Optical luminosities were evaluated at a common time of 11 hrs rest-frame, whereas X-ray luminosities were calculated at a rest-frame frequency of 1 keV and at a common rest-frame time of 2.2 d to match our coverage of GRB\,260310A. 
To bring GRB\,260310A into agreement with the prompt-afterglow correlations, the burst would require an isotropic-equivalent gamma-ray energy of $\gtrsim$10$^{52}$ erg,  nearly two orders of magnitude larger than the observed value. 
\added{
Observational biases affect the discovery of low-luminosity bursts (low $E_{\gamma,\rm iso}$) with faint afterglows (low $L_X$ and $L_{\rm opt}$), but they cannot explain why the right part of the diagram, corresponding to bright afterglows, remains sparsely populated. This suggests that GRB\,260310A occupies an unusual region of the parameter space, where standard on-axis GRBs are rarely found.
}

Moreover, as shown in Fig.~\ref{fig:gbmlc}c, the GRB prompt emission properties deviate from the locus of long GRBs in the Amati plane \citep{Amati2002,Amati_2006}. Its low isotropic-equivalent energy -- near the bottom of the observed distribution -- combined with a standard peak energy places it close to the region typically occupied by short GRBs, initially complicating its classification. However, the spectroscopic identification of an associated supernova unambiguously establishes a massive-star progenitor. We therefore interpret its inconsistency with the Amati relation as a viewing-angle effect, where a modest offset $\theta_v$ from the jet axis reduces the Doppler boosting, significantly lowering the prompt $\gamma$-ray output ($E_{\gamma, \rm iso} \propto \theta_v^{-6}$) while only mildly shifting the spectral peak to lower energies ($E_{\rm p} \propto \theta_v^{-2}$; \citealt{granot2002}). 

Several GRBs have been classified as candidate off-axis events on the basis of their low isotropic-equivalent energies and spectral properties. GRB\,980425, one of the closest GRBs ever detected ($z=0.0085$), 
was associated with SN\,1998bw and found to be less luminous by several orders 
of magnitude compared to the cosmological GRB population \citep{Pian_2000, 
Amati_2006, Zhang_2009}. \citet{Yamazaki2003} suggested that its 
soft spectrum and low $E_{\rm iso}$ could be reconciled with a standard long 
GRB viewed at a large angle from the jet axis. A similar suggestion was 
proposed for GRB\,031203 \citep{RamirezRuiz2005, 2005NCimC..28..351U, 
Zhang_2009}, another low-redshift ($z=0.11$), SN-associated burst whose 
prompt emission properties place it as an outlier of the Amati relation. 
GRB\,061021 and GRB\,161219B, while less extreme in their energetics, also 
show spectral and energetic properties broadly consistent with mildly off-axis 
viewing geometries \citep{Nava_2012, refId0}. In addition, GRB\,171205A 
($z=0.037$) was identified as a structured or two-component jet system, where 
the low prompt energy output and the late-time afterglow evolution were 
interpreted as the signature of a jet core observed at a modest offset 
\citep{Delia2018, Izzo2019}. These candidate off-axis events define an empirical locus in the  $E_{\rm p}$--$E_{\rm iso}$ plane that is systematically offset from the standard Amati relation \citep{Xu2023}.

An off-axis jet for GRB\,260310A naturally explains why the prompt $\gamma$-ray and early afterglow emission are suppressed while a comparatively bright afterglow is visible at later times, when the jet core has come into view. A viewing angle $(\theta_v - \theta_c)$\,$\approx$2-3 $\theta_c$, where $\theta_c$ is the half-opening angle of the jet core, would be sufficient to reconcile the observed prompt emission properties with the Amati relation, while keeping the inferred energetics within the range of the long GRB population ($\approx$\,10$^{52}-10^{53}$\,erg).
A value as high as $(\theta_v-\theta_c)\approx$6 $\theta_c$ would instead push the total isotropic-equivalent gamma-ray energy above 
10$^{55}$ erg, comparable if not higher than GRB 221009A \citep{Lesage2023,2023SciA....9I1405O}.

\begin{figure*}
    \centering
    \includegraphics[width=1\linewidth]{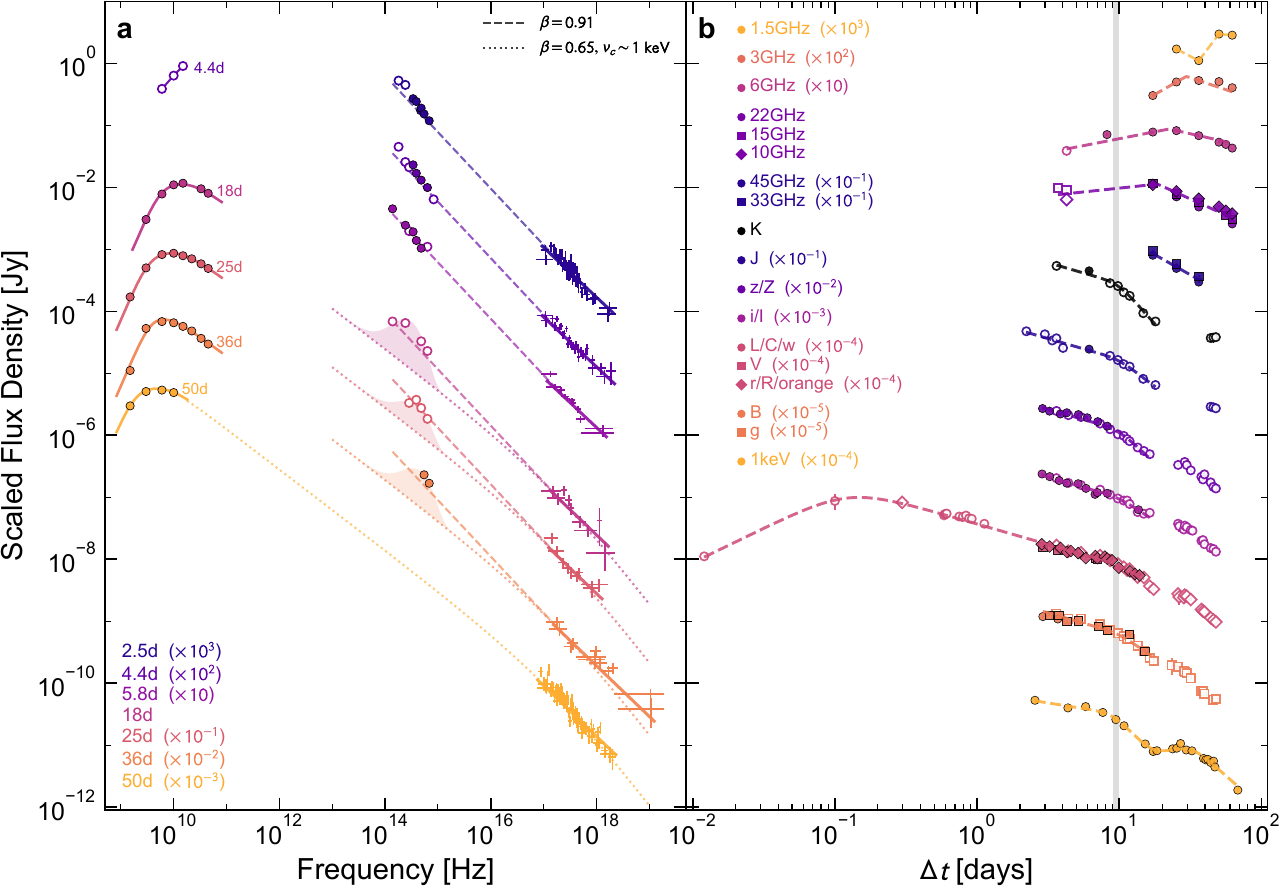}
    \caption{\added{Multi-wavelength observations of GRB\,260310A. Filled markers are data obtained in this work, with complementary public data shown as open markers. Data are rescaled for display.
    \textbf{a.} Spectral energy distribution (SED) from 2.5 d to 50 d. 
    The dashed lines show the extrapolation from 1~keV to the $K$ band assuming a power-law spectrum with $\beta=0.91$. This model can bridge the optical/nIR and X-ray bands at all epochs. However, such an extrapolation would leave little room for the SN emission that instead dominates the optical bands.
    The dotted lines indicate a smoothly broken synchrotron model with $\beta\sim0.65$ and  $\nu_c\sim1$ ~keV, which allows for an additional SN component (shaded blackbody models). 
    \textbf{b.} Multi-wavelength light curves ranging from X-ray (bottom), through optical and NIR, to radio frequencies (top). 
    The vertical band marks the observed temporal break in the X-ray and optical/NIR data.}
    }
    \label{fig:SEDLC}
\end{figure*}

\subsection{Basic afterglow constraints}

\subsubsection{Evidence for a bright reverse shock}\label{sec:rs}

The early X-ray and optical afterglows, from $T_0+$2 d to $T_0+$14 d, exhibit a similar (although not consistent within the errors) temporal behavior characterized by a shallow decay with slope $\approx0.4-0.6$, a temporal break at $\approx$9 d, and a steep decay $\alpha \approx 2$ \added{(see Fig.~\ref{fig:opt} and Fig.~\ref{fig:SEDLC}b)}. Their spectral slopes, 
$\beta_X$=0.80$\pm$0.05 and $\beta_o$=0.91$^{+0.06}_{-0.09}$, are consistent with each other, and their flux densities align on the same power-law segment, with no evidence for a spectral break between the two bands \added{(see Fig.~\ref{fig:SEDLC}a)}.

Taken together, these results support the interpretation that the early optical and X-ray emission arises from the same spectral component and belongs to the same synchrotron segment.  In the context of the standard synchrotron afterglow model \citep{sari1998,wijers1999}, where the observed radiation arises from a population of shock-accelerated electrons following a power-law energy distribution with index $p$, the optical-to-X-ray spectrum would be consistent either with the regime between the synchrotron characteristic frequency $\nu_m$ and the cooling frequency $\nu_c$ ($\nu_m < \nu_o < \nu_X < \nu_c$) for $p\approx2.8$ or with the regime $\nu_m < \nu_c < \nu_o < \nu_X$ for $p\approx1.8$.

Neither of these values of $p$ is consistent with the flatter electron index, $p\approx2-2.3$, inferred from the radio spectrum (Fig.~\ref{fig:radioSED}), which already suggests the presence of an additional emission component peaking at lower frequencies. We therefore consider a standard two-component scenario, as also discussed in \citet{Christy2026}, in which the high-energy emission traces synchrotron radiation from the forward shock (FS) propagating into the external medium, while the low-energy spectrum is dominated by emission from a reverse shock (RS) traveling back into the ejecta \citep{meszaros1997,Sari1999,kobayashi2000,zhang2003,Gomboc2008}.

Our initial hypothesis is that, throughout the entire temporal evolution, the non-thermal optical and X-ray emission belong to the same power-law segment of the FS component. However, this interpretation breaks down at late times. As the X-ray emission undergoes a rebrightening without any significant spectral evolution (Sect.~\ref{sec:xray}), extrapolating the X-ray spectrum into the optical band would overpredict the observed optical flux, as illustrated in Fig.~\ref{fig:opt} \added{and Fig.~\ref{fig:SEDLC}a}. 
Furthermore, optical spectroscopy (see inset of Fig.~\ref{fig:opt}) provides clear evidence for the emergence of a SN component
\citep{OConnor2026} rather than a non-thermal power-law spectrum.
Based on this evidence, we infer that a spectral break between the X-ray and optical bands must be present in order to reconcile all the observations.  
This implies that FS radiation is not the dominant component at optical and nIR wavelengths, as initially postulated. Instead, the observed emission is likely composed of a combination of FS+RS radiation at early times, gradually transitioning to a mixture of FS+SN emission at later epochs.
A RS contribution helps explain the different X-ray and optical temporal slopes and the achromatic break at $\approx$9 d, which we interpret as a jet break. The RS polarized light would be diluted by the FS component, making this scenario consistent with the optical polarization limits (Sect.~\ref{sec:opol}).

Given the complex afterglow evolution caused by the interplay of multiple components, standard closure relations for uniform or structured jets 
\citep[e.g.][]{Gao2013,Ryan2020} cannot be successfully applied
and it becomes challenging to map the transition fro
nbvhm RS-dominated to FS-dominated emission. Our polarimetric measurement helps in this respect. 
At $T_0+55$ d our radio observations probe the optically-thin spectrum and derive a low degree of linear polarization ($\approx$1.7\%), consistent with typical FS values  \citep{Greiner2003, Covino2016,Mandarakas2023} and lower than the $\approx$3.2\% polarization fraction derived by \citet{Christy2026} for the optically-thin RS-dominated spectrum. 
Therefore, we conclude that at this epoch, FS radiation is the dominant non-thermal component, from radio to X-rays. 
The radio-to-X-ray spectral index is $\beta_{RX} \approx$0.65, which implies an electrons' spectral index $p\approx$2.3. 
To explain the observed X-ray flux and spectral index $\beta_X\approx0.8$,
we estimate that the cooling frequency $\nu_c$ falls within the  X-ray band, where its intrinsically smooth break prevents any sharp spectral transition within the X-ray band \added{(see Fig.~\ref{fig:SEDLC}a)}.

\subsubsection{A jet viewed off-axis}\label{sec:view}

The early optical afterglow of GRB\,260310A is characterized by an initial rising phase, brightening from $\approx18.8$ mag at 17 min to $\approx16.7$ mag at 0.3 d, before transitioning to a shallow decay with $\alpha_O \approx 0.6$. 
For highly relativistic outflows with Lorentz factors $\Gamma \gg 100$ viewed on-axis, the afterglow peak is expected to occur within minutes of the burst \citep{Molinari2007}, and is therefore rarely captured by follow-up observations \citep{Panaitescu2008,Rykoff2009,Melandri2014,Dainotti2024}. 
A delayed peak may instead indicate either a moderately relativistic outflow, implying a relatively low initial Lorentz factor \citep[e.g.,][]{Lipunov2022,Ho2022}, or a geometric effect, such as an observer viewing angle offset from the jet axis \citep[e.g.,][]{Guidorzi2009, Kruhler2009, troja2017,Troja2018b}.

Adopting a luminosity of $L \sim 10^{49}$ erg s$^{-1}$ and a variability timescale of $\Delta t_{\min} \lesssim 5$ s, a Lorentz factor of $\Gamma \gtrsim 10$ would be sufficient to explain the observed gamma-ray spectrum. This value is consistent with a moderately baryon-loaded (``dirty") fireball \citep{Paczynski1998,Dermer1999}, indicating that compactness arguments alone cannot discriminate between low-Lorentz-factor outflows and geometric effects in this event.

\added{Polarization does not provide a decisive diagnostic either. Significant linear polarization is expected from geometric asymmetry, such as an off-axis viewing configuration \citep{Rossi2004,Birenbaum2026}. However, the degree of polarization and its temporal evolution depend sensitively on the jet structure, magnetic-field geometry, reverse-shock contribution, and spectral regime. In the case of GRB\,260310A, the combination of FS and RS components, along with limited temporal sampling of the polarimetric measurements and the depolarization by synchrotron self-absorption at low radio frequencies \citep{Jones1977}, prevents a unique interpretation.}

\added{Additional insights come from the Amati diagram: 
the burst exhibits a peak energy that is harder than the average population of long GRBs with similar energetics. Baryon-loaded outflows are instead expected to produce softer spectra, as the reduced Lorentz factor lowers the characteristic photon energies \citep{Zhang_Mezaros_2002, Sakamoto2005}. 
The relatively hard spectrum of GRB\,260310A tends to disfavor a dirty fireball, whereas it remains consistent with a classical GRB seen off-axis (see also Sect.~\ref{sec:context}).}

\added{The temporal slope of the afterglow light curve will be shaped by (pre-)deceleration dynamics, jet geometry, or a combination of both. An optical FS light curve observed between $\nu_m$ and $\nu_c$ (as argued in sect~\ref{sec:rs}) will have an asymptotic rising slope $t^3$ prior to deceleration. This both holds for jets observed fully on-axis and off-axis, assuming the line-of-sight is included within the beaming cone of flow coasting at constant Lorentz factor $\Gamma (\theta)$, i.e. $\theta_v - \theta_{\rm w} < 1 / \Gamma (\theta_{\rm w}$). An off-axis observer, however, would see a more extended period of turnover across the peak, leading to an apparently shallower rise in case the full asymptotic rise stage is not captured by observation.
}

\added{An observer far off-axis, with $\theta_v - \theta_{\rm w} > 1 / \Gamma (\theta_{\rm w}$), will see an even steeper rise following $t^{(15-3p)/2} \sim t^{4.1}$ (assuming $p \sim 2.3$), once the jet has begun to decelerate, regardless of jet structure.
After the jet truncation angle $\theta_{\rm w}$ has entered into the field of view, a rising temporal slope $\alpha < 0$ is still possible for a smoothly structured jet (e.g. a Gaussian or power-law jet, rather than a two-component jet), if the observer angle $\theta_v$ is sufficiently larger than the core angle $\theta_c$ \citep{Ryan2020}.}

\added{However, if the jet is viewed off-axis, but the observer angle $\theta_\nu$ follows $\theta_c < \theta_\nu < \theta_w$, the temporal slope due to jet structure will not be negative but instead lead to a shallower decline than predicted by an on-axis model (tophat or structured). In this interpretation, the pre-peak light curve for GRB 260310A can be attributed to pre-deceleration coasting, whereas the relatively shallow post-peak slope (compared to top-hat models) reflects emission from higher energy angles of the flow closer to the jet axis gradually entering in to the line of sight. 
}
\added{
The turnover at nine days is then a consequence of the outer truncation angle of the jet coming into view, i.e. when $1 / \Gamma (\theta_{\rm w}) \sim \theta_{\rm w} - \theta_{\nu} $. The presence of a late-time rebrightening 
is consistent with the hypothesis that GRB 260310A is seen off-axis, as we will discuss below.}

\begin{figure}
    \centering
    \includegraphics[width=\linewidth]{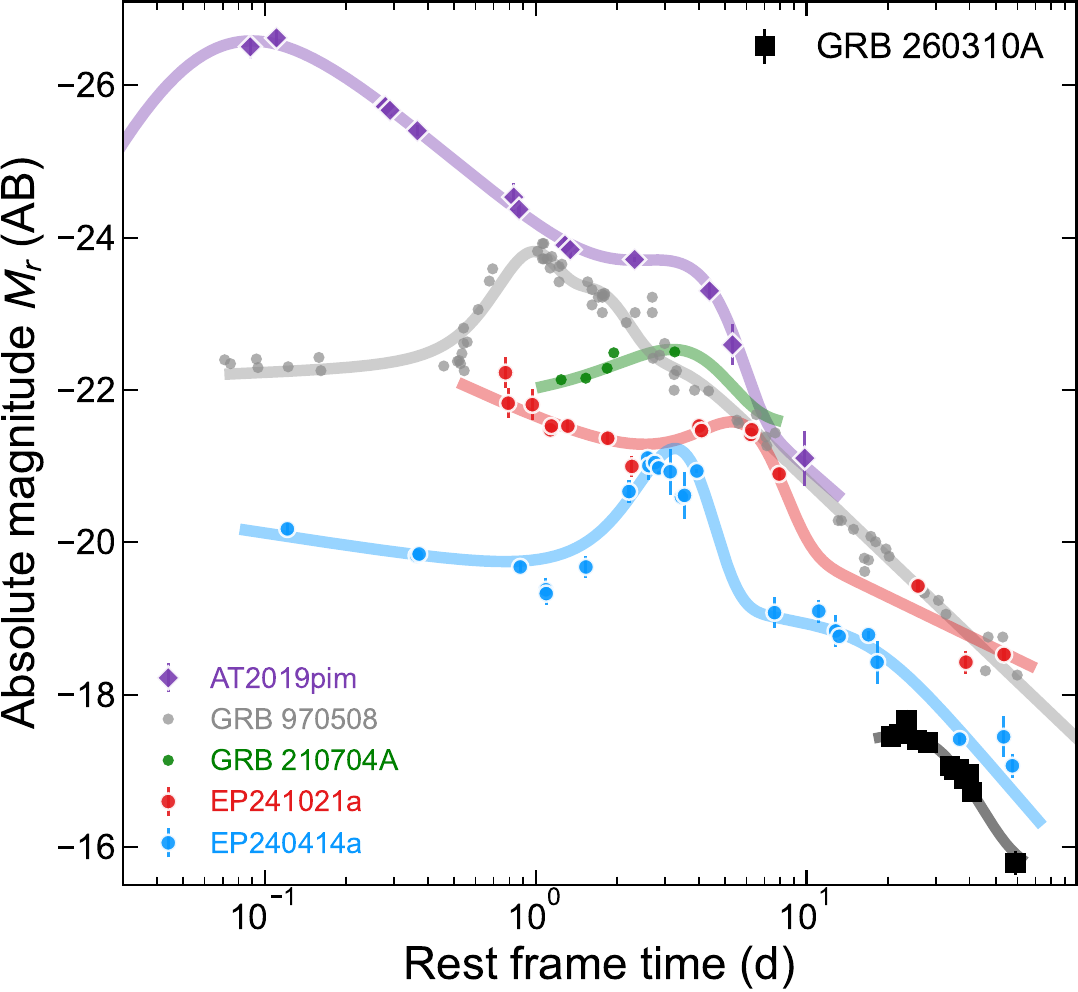}
    \caption{The rebrightening of GRB\,260310A (black squares), scaled from X-rays to optical assuming $\nu_c\approx$\,0.1 keV and a power-law index $\beta$\,$\sim$0.6. A sample of late-time afterglow rebrightenings, including AT2019pim \citep{Perley2025}, GRB 970508 \citep{Galama970508}, GRB 210704A (assuming $z\approx$2.3; \citealt{Becerra2023}), EP241021a \citep{Busmann2025}, and EP240414a \citep{Sun2025}, is shown for comparison. 
    }
    \label{fig:lcbump}
\end{figure}

\subsubsection{Origin of the late-time afterglow rebrightening }
\label{sec:bump}

Long-lasting X-ray emission has been observed in other nearby GRBs \citep{Kouveliotou2004,Margutti2013}, typically with spectra softer than those of standard afterglows and therefore often interpreted as arising from an additional emission component. In GRB\,260310A, inverse-Compton (IC) scattering of SN optical photons \citep{Chevalier2006} would be an intriguing possibility, since the X-ray bump occurs on a timescale comparable to the expected SN peak \citep{OConnor2026}. 
In such a scenario, if the seed blackbody photons are
scattered by the bulk of the shocked electrons, with
$\gamma_e \sim \gamma_m$, the characteristic upscattered frequency would be 
$\nu_{\rm BB,IC} \sim \gamma_m^2 \nu_{\rm BB}$,
where $\gamma_m$ is the minimum Lorentz factor of the accelerated electrons.
This can be recast as: 
\begin{eqnarray}
\nu_{\rm BB, IC} & \sim & 1.9 \times 10^{19} \left( \frac{\epsilon_B}{10^{-3}} \right)^{-\frac{1}{2}} \left( \frac{\Gamma}{10} \right)^{-2} \times \nonumber \\
& & \left( \frac{\nu_m}{10^{10} \textrm{ Hz}}\right) \left( \frac{\nu_{\rm BB}}{10^{15} \textrm{ Hz}}\right) \left( \frac{n_{\rm ext}}{ 1 {\rm~cm}^{-3}} \right)^{-\frac{1}{2}} \textrm{ Hz}
\end{eqnarray}
Thus, for typical parameters, the IC peak is expected to lie at or above
the X-ray band, rather than producing a featureless X-ray
component with little spectral evolution.
This, together with the observed luminosity, $L_X \approx 10^{43}\,{\rm erg\,s^{-1}}$, disfavors an IC origin of the X-ray emission. 
Instead, the stable X-ray spectral index and the radio SED indicate that synchrotron emission from the FS remains the dominant component in the X-ray band.

As shown in Fig.~\ref{fig:lcbump}, afterglow rebrightenings on rest-frame timescales of $\approx1-10$ d are sporadically observed in GRBs and fast X-ray transients (FXTs) and are commonly interpreted as episodes of late energy injection refreshing the blast wave \citep{rees1998}. In this context, GRB\,260310A shows a rebrightening that evolves on somewhat longer timescales, $\approx20$--30 d, and is fainter than most events in the comparison sample. However, the observed diversity of late-time afterglows is poorly characterized and strongly affected by selection biases. 
In the optical band, the SN onset would mask most of these late-time brightenings. More in general, follow-up observations are typically interrupted once the afterglow becomes faint, and only events considered especially interesting, such as GRB\,260310A, are monitored with sufficient cadence and depth to reveal their late-time structures.
For instance, a query of the full \textit{Swift}  X-ray afterglow archive \citep{Evans2009} shows that only $\approx$1\% of the sample, 13 out of more than 1300 long GRBs, have X-ray coverage extending beyond 15 days. 
An inspection of the optical catalog of \citet{Dainotti2024} returns similar results.  
The apparent rarity of late-time afterglow rebrightenings may therefore reflect, at least in part, the limited temporal and spectral coverage of most monitoring campaigns rather than their intrinsic absence.

In the case of GRB\,260310A, its off-axis geometry may offer an alternative route to explain the late-time X-ray rebrightening. 
A gradual afterglow rebrightening can be expected when a structured (or two-component) jet is viewed off-axis \citep[e.g.][]{Beniamini2022}.  
In this framework, the early optical afterglow is dominated by the wider and less energetic component,
\added{
which exhibits a jet break at $\approx$9\,d as 
the nearest outer jet edge comes into view (i.e. $1 / \Gamma (\theta_{\rm w}) \sim \theta_{\rm w} - \theta_{\nu} $). For an off-axis observer viewing a two-component jet, this might well occur before the inner component enters the line of sight.
}
The later X-ray bump is \added{then} attributed to \added{this}
narrower and more energetic jet core, which becomes visible only after sufficient deceleration reduces relativistic beaming and brings its emission into the observer's line of sight. Thus, the two afterglow peaks, at $\approx$0.1\,d and $\approx$25\,d,  could be interpreted as the emergence of distinct angular components of the outflow.

A two-component jet structure has often been invoked to explain the evolution of the brightest GRBs and their afterglows, whose behavior deviates from canonical closure relations. A remarkable example is the recent GRB\,221009A, for which the extreme brightness of the prompt emission and the absence of a jet break in the afterglow data would imply an energy crisis unless a complex jet structure is taken into account \citep{2023SciA....9I1405O,Sato2025}. A similar geometry has also been proposed to explain the origin of some gamma-ray dark fast X-ray transients characterized by late-time rebrightening \citep{Busmann2025,Yadav2025,Gianfagna2025, Sun2025, Zheng2025}. 
In the case of GRB\,260310A, a narrow gamma-ray emitting core surrounded by broader and less energetic wings and seen off-axis could account for the peculiar X-ray afterglow evolution, the delayed afterglow onset and the burst location in the Amati plane.

\subsection{Afterglow modeling}
\label{sec:afterglow_modeling}

We fit the broadband data using 
\texttt{VegasAfterglow} \citep{Wang2026_VegasAfterglow},
which allows us to include both FS and RS components. 
The modeling is based on the standard external shock synchrotron emission from a relativistic jet interacting with the circumburst medium \citep{meszaros1997,sari1998, wijers1999, Granot2002breaks, granot2002, vanEerten2013, Ryan2020}. 

The intrinsic energetics and dynamics are governed by the isotropic-equivalent kinetic energy $E_{\text{k,iso}}$ and the initial Lorentz factor $\Gamma_0$, while the microphysical properties are described by the fractions of shock energy partitioned into electrons $\epsilon_e$ and magnetic fields $\epsilon_B$, along with the power-law index of the electron energy distribution $p$. 
The jet profile is characterized by the core half-opening angle $\theta_c$, supplemented by the energy distribution index $k_e$ and the Lorentz factor distribution index $k_g$ for the power-law structured model.
The observer’s viewing angle $\theta_{\text{v}}$ is treated as a free parameter to account for off-axis effects. 
In all cases, the fraction of accelerated electrons is fixed at $\xi_{e}=1$.  We tested different jet geometries, including uniform, Gaussian, and power-law profiles, as well as a two-component jet. 
The parameter space was explored using the Python package \texttt{pymultinest} \citep{Buchner2014}.
 
We find that a simple FS model from an off-axis structured jet provides a reasonable description of the broadband light curves, from radio to X-rays (Fig.~\ref{fig:gauss_ism_fs} left panel).  By using a flat spectral index $p\approx2$ (Fig.~\ref{fig:gauss_ism_fs} right panel) and a uniform density ($n_{\text{ISM}}$) environment,  the model can reproduce most of our dataset up to $\approx$ 20 d, when the X-ray afterglow brightens. 
This model has $\nu_a \approx 6$ GHz, $\nu_m\approx 12$ GHz, and  $\nu_c\approx 3\times 10^{14}$ Hz at 17 d.
However, this solution is driven by our limited coverage of the hard X-ray and mm range, 
which are systematically underpredicted by the FS model, the 
similar behavior of the X-ray and optical/nIR light curves (Sect.~\ref{sec:rs}), and it does not account for the SN contribution in the optical range. 
Therefore, whereas this model roughly describes the observed light curves and spectra, it does not fully reproduce the physics of the explosion. 

No combination of RS + FS provides a better fit for either a uniform or a wind-like density profile. This is not unexpected, 
as the empirical description of the radio SEDs (Sect.~\ref{sec:radiosed}) already revealed that the spectral evolution deviates from standard predictions,
and suggests that a more complex scenario is required, potentially involving mechanisms such as energy injection and refreshed shocks.
These could also be used to explain the late-time bump of the X-ray light curve. 

As discussed in Sect.~\ref{sec:bump}, we also test whether the late-time emission could be interpreted as the emergence of a second narrow-jet seen off-axis. We model the X-ray data and the radio data after $T_0+50$\,d with a simple FS model from a two-component jet expanding into a uniform medium.
This model (Fig.~\ref{fig:2com}) can reproduce the late-time bump with a narrow, highly relativistic and energetic jet core, with $\theta_c \approx 0.8^\circ$, $\Gamma \approx 150$, and $E_{\rm K,iso} \approx 2 \times 10^{54}$ erg, viewed at an angle $\theta_v \approx 10^\circ$. The core is surrounded by a broader, slower, and less energetic jet component, with $\theta_w \approx 16^\circ$, $\Gamma \approx 15$, and $E_{\rm K,iso} \approx 5 \times 10^{52}$ erg, which accounts for the early X-ray emission.

\section{Summary}

We presented multi-wavelength observations of the nearby ($z\sim0.153$) GRB\,260310A and its extremely bright and long-lived afterglow, 
placing it among the top $\sim 1\%$ of the GRB afterglow population.
The burst is characterized by a modest gamma-ray energy release
paired with a relatively hard peak energy, which classifies it as an outlier of the Amati relation. 
We showed that viewing effects solve the apparent tension in the burst's properties. The slow rise and delayed onset of its optical afterglow support the scenario of a jet seen off-axis.

Our analysis of the broadband spectrum favors the presence of two emission components: a bright reverse shock dominating the radio and extending to the optical band, and a forward shock contributing to the optical and dominating in the X-rays. In support of this picture, we infer a decreased degree of linear polarization, from $\Pi \sim 3.2\%$ to $\Pi \sim 1.7\%$, mapping the transition from reverse- to forward-shock dominated emission. 
However, as often occurs for bright GRB afterglows, none of our standard reverse-shock plus forward-shock models, testing different jet structures and density profiles, can successfully reproduce the full broadband dataset and its temporal evolution, suggesting a departure from the simplest afterglow prescriptions.

Our long-term X-ray monitoring of the X-ray afterglow identifies a peculiar late-time rebrightening at $\approx$20-30 d, after a phase of rapid decay. This behavior is difficult to accommodate within a single forward-shock model and may instead indicate a late episode of energy injection that refreshes the shock or additional structure in the outflow, such as a two-component jet.
\added{A low-velocity jet seen (approximately) on-axis would not be able to combine the early peak and late rebrightening in a single jet model, whereas an off-axis two-component jet can accommodate both these peaks.}

\begin{acknowledgments}
This work is supported by the European Research Council through the Consolidator grant BHianca (grant agreement ID~101002761). 

AJCT acknowledges support from the Spanish Ministry project PID2023-151905OB-I00 and Junta de Andaluc\'ia grant P20\_010168. AJCT expresses his sincere thanks to the staff of the San Pedro M\'artir Observatory in M\'exico and IHSM/UMA-CSIC La Mayora in Spain for their assistance in the operation of the 0.6m Javier Gorosabel and TELMA telescopes at the BOOTES-5 and BOOTES-2 stations, respectively. 

Data were partially collected with the 0.9m and 1.5m telescopes at the Observatorio de Sierra Nevada (OSN) operated by the Instituto de Astrofísica de Andalucía (IAA-CSIC), and based on observations collected at the Centro Astron\'omico Hispano en Andaluc\'ia (CAHA) at Calar Alto (proposal 26A-2.2-026), operated jointly by Junta de Andaluc\'ia and Consejo Superior de Investigaciones Cient\'ificas (IAA-CSIC).

This work is based on the data obtained with Einstein Probe, a space mission supported by the Strategic Priority Program on Space Science of Chinese Academy of Sciences, in collaboration with the European Space Agency, the Max-Planck-Institute for extraterrestrial Physics (Germany), and the Centre National d'Études Spatiales (France).

We acknowledge the support from the LBT-Italian Coordination Facility for the execution of observations, data distribution and reduction. Reduction and analysis by LBT Imaging Data Center at INAF-OARoma.
The LBT is an international collaboration among institutions in Italy, the United States and Germany. LBT Corporation Members are: Istituto Nazionale di Astrofisica, Italy; The University of Arizona on behalf of the Arizona Board of Regents; LBT Beteiligungsgesellschaft, Germany, representing the Max-Planck Society, The Leibniz Institute for Astrophysics Potsdam, and Heidelberg University; The Ohio State University, representing OSU, University of Notre Dame, University of Minnesota and University of Virginia. Observations have benefited from the use of ALTA Center (alta.arcetri.inaf.it) forecasts performed with the Astro-Meso-Nh model. Initialization data of the ALTA automatic forecast system come from the General Circulation Model (HRES) of the European Centre for Medium Range Weather Forecasts.

\added{This research has made use of data obtained from the Chandra Data Archive provided by the Chandra X-ray Center (CXC). This paper employs a list of Chandra datasets, obtained by the Chandra X-ray Observatory, contained in~\dataset[DOI: 10.25574/cdc.620]{https://doi.org/10.25574/cdc.620}.}

Sky-background data is provided by \texttt{ATRAN} \citep{Lord1992} and Gemini Observatory.

\end{acknowledgments}

\software{
\texttt{Astropy} \citep{astropy:2013, astropy:2018, astropy:2022}, 
\texttt{Matplotlib} \citep{matplotlib}, 
\texttt{SciPy} \citep{Virtanen2020}, 
\texttt{NumPy} \citep{numpy},
\texttt{pandas} \citep{pandas},
\texttt{Source Extractor} \citep{Bertin1996}, 
\texttt{SFFT} \citep{Hu2022}, 
\texttt{SAOImageDS9} \citep{Joye2003}, 
\texttt{dustmaps} \citep{dustmaps},
\texttt{dust\_extinction} \citep{dust_extinction},
\texttt{VegasAfterglow} \citep{Wang2026_VegasAfterglow},
\added{\texttt{HEASoft} \citep{heasoft}}
}

\appendix
\renewcommand{\thetable}{A\arabic{table}}
\renewcommand{\thefigure}{A\arabic{figure}}
\setcounter{table}{0}
\setcounter{figure}{0}

\section{Details of Optical Observations}
\label{sec:optical_details}

\textit{BOOTES}: Two robotic telescopes, BOOTES-2 (Spain) and BOOTES-5 (M\'exico) were used and observations were carried out in the Sloan $g^\prime$, $r^\prime$, $i^\prime$  and UKIRT $Z$ filters, covering from $T_0 + 3$~d to $T_0 + 15$~d, with the seeing of $\sim1.8''-4.6''$.
We employed the saccadic Fast Fourier Transform (SFFT) algorithm \citep{Hu2022} to isolate the transient emission from the host galaxy background. 

\textit{OSN}: Observations were conducted using the 1.5~m telescope (T150) and the 0.9~m telescope (T90) in the Johnson-Cousins filters $B$, $V$, $R$, $I$ ($2\times2$ binning) starting from $\sim T_0 + 3$\,d. The average seeing was 1.5\arcsec~for T150 and 2.5\arcsec~for T90.
Image reduction and photometry were performed using \texttt{MaxIm DL} v6.13.
Photometric calibration was performed using Pan-STARRS DR1 \citep{Chambers2016} reference stars, converted to the $BVRI$ system with the Lupton (2005)\footnote{\url{https://www.sdss4.org/dr12/algorithms/sdssubvritransform/\#Lupton2005}.} color-transformation equations.

\textit{Ond\v{r}ejov}: Observations at the Ond\v{r}ejov Observatory (Czech Republic) were conducted using three telescopes. The 2 $\times$ 20 cm Small Binocular Telescope \citep[SBT;][]{2019AN....340..633S} was employed with the Sloan $g^\prime$, $r^\prime$, and $i^\prime$ filters. 
Concurrent high-cadence monitoring in the Sloan $r^\prime$ band (8 epochs between $T_0 + 4$~d and $T_0 + 14$~d) 
were provided by the 0.5 m robotic telescope \citep[D50;][]{2023CoSka..53d..49S}. The 2.0 m Perek telescope (O2m) contributed to the campaign, utilizing the Sloan $g^\prime$ filter at the beginning of the observations ($T_0 + 4$~d).
The typical seeing was $\sim$3\arcsec~throughout the observations at Ond\v{r}ejov. Image processing consisted of per-frame dark and flat-field calibration, followed by inverse-variance-weighted coaddition via the {\tt pyrt} pipeline \citep{2023CoSka..53d.127J}; SBT frames were additionally drizzled prior to coaddition owing to the undersampled plate scale of that instrument. Host-galaxy contamination was removed by subtraction of a PanSTARRS~DR1 template using {\sc hotpants} \citep{2015ascl.soft04004B}.

\textit{LBT}: We used the LBT Utility Camera in the Infrared (LUCI), with $J$ and $K$ filters, and the Large Binocular Camera (LBC), with \textit{r} and \textit{z} filters, instruments mounted on LBT under program ID IT-2025B-047 (PI: Troja). 
LUCI observed at $T_0 + 6.1$ d with an average air mass of 1.4 and seeing of 0.7\,". 
LBC observed at $T_0 + 8.2$ days with an average air mass of 1.3 and mean seeing of 1.1\,". 
We subtracted the host contamination for $r$ and $z$ filters against LS archival images using the SFFT.

The observed magnitudes, uncorrected for Galactic extinction, are listed in Tab.~\ref{tab:obslog}.

\textit{XMM-Newton Optical Monitor}:
The XMM-Newton Optical Monitor (OM) conducted simultaneous observations using the \textit{U}, \textit{V}, \textit{B}, \textit{UVW1}, \textit{UVW2}, and \textit{UVM2} filters while EPIC was acquiring X-ray data. The exposure times were 8.8 ks for the \textit{UVW2} filter and 4.4 ks for the remaining filters. Data were processed using SAS \texttt{omichain}. 
Whereas optical images are dominated by the galaxy's light, 
in the ultraviolet filters a faint source is detected at the GRB position with $m_{UVW2} =$ 21.7 $\pm$0.4 AB mag (corrected for extinction along the sightline). 
However, based on a single epoch we cannot assess
whether the source is transient or associated with an underlying star-forming knot \citep{OConnor2026}. \bigskip \\

\begin{figure*}
    \includegraphics[width=\linewidth]{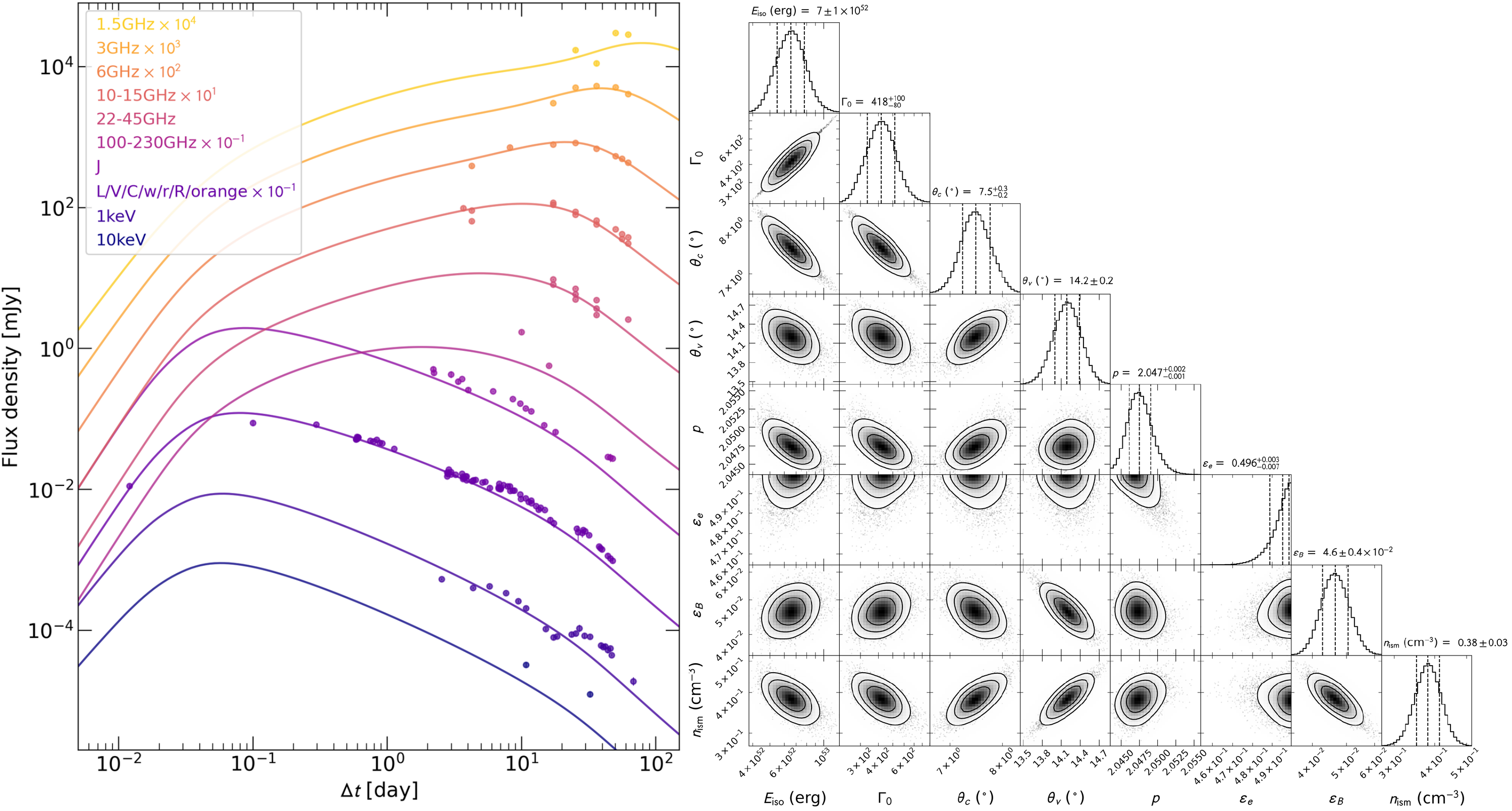}
    \caption{Left: Multi-wavelength light curves of GRB\,260310A together with the best-fit Gaussian jet  FS model in an ISM environment.
    Right: Posterior distributions of the model parameters. The median values and corresponding 1$\sigma$ uncertainties are indicated above the one-dimensional marginalized histograms.
    }
    \label{fig:gauss_ism_fs}
\end{figure*}

\begin{figure*}
    \includegraphics[width=\linewidth]{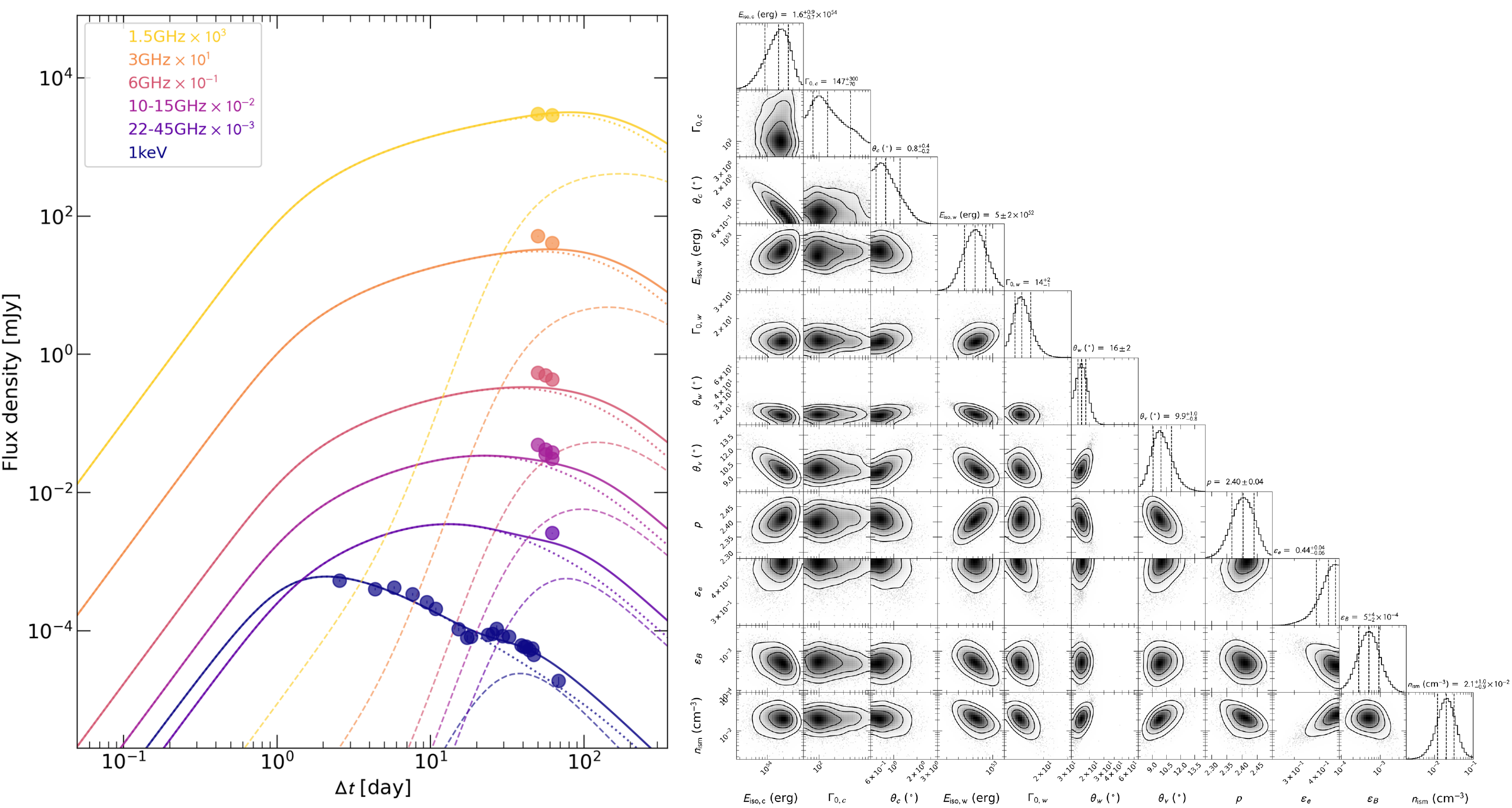}
    \caption{Left: Multi-wavelength light curves of GRB\,260310A together with the best-fit forward shock emission from the two-component jet in an ISM environment, adopted to describe the X-ray and late radio data. In this scenario, the early optical and radio emission are dominated by a RS and are treated as upper limits in the fit. 
    Right: Posterior distributions of the model parameters. The median values and corresponding 1$\sigma$ uncertainties are indicated above the one-dimensional marginalized histograms.
    }
    \label{fig:2com}
\end{figure*}

\begin{ThreePartTable}
    \begin{TableNotes}[para]
        \footnotesize
        Notes: $\Delta t$ represents the mid-time of the observation relative to the trigger time $T_0$. 
        An error of 0.3 mag was associated to the DDOTI measurement to account for uncertainties in the photometric calibration. 
        References: 1. \citealt{GCNGOTO}, 2. \citealt{GCNDDOTI}, 
        3. J. Tonry et al. 2026 (\url{https://www.wis-tns.org/object/2026fgk}), 4. \citealt{GCN43974}, 5. \citealt{GCN43996}, 6. \citealt{GCN43993}, 7. \citealt{GCN44004}, 8. \citealt{Rhodes2026}, 9. \citealt{Giarratana2026}, 10. \citealt{Ho2026a}, 11. \citealt{Ho2026b}.
    \end{TableNotes}
\setlength{\LTcapwidth}{\textwidth}
\begin{longtable}{@{\extracolsep{\fill}}lccccc}
\caption{Log of X-ray, optical, NIR, and radio observations of GRB 260310A. \textit{The full dataset will be available in a machine-readable format.}} \label{tab:obslog}
\endfirsthead
\multicolumn{6}{c}{\tablename\ \thetable{} -- \textit{Continued from previous page}} \\
\midrule
\endhead
\midrule
\multicolumn{6}{r}{\textit{Next page}} \\
\endfoot
\bottomrule
\insertTableNotes
\endlastfoot

\hline
\multicolumn{6}{l}{\textbf{X-Ray}}\\
Date& $\Delta t$& Telescope & Exposure &Flux & Flux density\\
(UT) & [d] &  & [ks] &[erg\,cm$^{-2}$\,s$^{-1}$] &[mJy]\\
\hline
2026-03-12 18:12:41 & 2.55 & EP/FXT-A+B & 3.3 &  $5.1\pm0.2\times 10^{-12}$ &  $5.3\pm0.2\times 10^{-4}$ \\
2026-03-14 13:23:51 & 4.35 & EP/FXT-A+B & 3.9 &  $3.9\pm0.1\times 10^{-12}$ &  $4.0\pm0.1\times 10^{-4}$ \\
2026-03-16 00:01:31 & 5.79 & EP/FXT-A+B & 3.0 &  $4.0\pm0.1\times 10^{-12}$ &  $4.2\pm0.2\times 10^{-4}$ \\
2026-03-17 20:44:25 & 7.66 & EP/FXT-A+B & 2.8 &  $3.3\pm0.1\times 10^{-12}$ &  $3.4\pm0.1\times 10^{-4}$ \\
2026-03-19 15:51:50 & 9.45 & EP/FXT-A+B & 2.8 &  $2.5\pm0.1\times 10^{-12}$ &  $2.6\pm0.1\times 10^{-4}$ \\
2026-03-21 00:01:09 & 10.8 & NuSTAR/FPM-A+B & 27.6 &  $2.8\pm0.1\times 10^{-12}$ (3--79 keV) &  $3.2\pm0.3\times 10^{-5}$ (10 keV) \\[-2pt]
2026-03-21 06:11:59 & 11.1 & EP/FXT-A+B & 2.7 &  $1.98\pm0.07\times 10^{-12}$ &  $2.1\pm0.1\times 10^{-4}$ \\
2026-03-25 10:00:09 & 15.2 & EP/FXT-A+B & 4.9 &  $10.0\pm0.5\times 10^{-13}$ &  $1.05\pm0.06\times 10^{-4}$ \\
2026-03-27 12:19:52 & 17.3 & EP/FXT-A+B & 2.9 &  $7.5\pm0.6\times 10^{-13}$ &  $7.9_{-0.6}^{+0.7}\times 10^{-5}$ \\
2026-03-28 13:52:35 & 18.4 & EP/FXT-A+B & 9.0 &  $7.8\pm0.4\times 10^{-13}$ &  $8.1\pm0.4\times 10^{-5}$ \\
2026-04-02 21:37:07 & 23.7 & EP/FXT-A+B & 3.3 &  $8.3\pm0.6\times 10^{-13}$ &  $8.7_{-0.6}^{+0.7}\times 10^{-5}$ \\
2026-04-04 13:32:46 & 25.4 & EP/FXT-A+B & 3.3 &  $8.5\pm0.6\times 10^{-13}$ &  $9.0_{-0.6}^{+0.7}\times 10^{-5}$ \\
2026-04-06 03:58:19 & 27.0 & EP/FXT-A+B & 1.0 &  $1.0\pm0.1\times 10^{-12}$ &  $1.1\pm0.1\times 10^{-4}$ \\
2026-04-08 18:17:16 & 29.6 & EP/FXT-A+B & 2.6 &  $8.0_{-0.7}^{+0.6}\times 10^{-13}$ &  $8.4\pm0.7\times 10^{-5}$ \\
2026-04-11 15:12:08 & 32.4 & EP/FXT-A+B & 4.2 &  $7.7\pm0.4\times 10^{-13}$ &  $8.1\pm0.5\times 10^{-5}$ \\[-2pt]
2026-04-11 16:53:39 & 32.5 & NuSTAR/FPM-A+B & 12.8 &  $1.07_{-0.09}^{+0.08}\times 10^{-12}$ (3--79 keV) &  $1.2\pm0.1\times 10^{-5}$ (10 keV) \\
2026-04-18 06:59:52 & 39.1 & EP/FXT-A+B & 2.7 &  $5.8_{-0.6}^{+0.5}\times 10^{-13}$ &  $6.1\pm0.6\times 10^{-5}$ \\
2026-04-19 18:54:26 & 40.6 & EP/FXT-A+B & 6.2 & \multirow{2}{*}{ $5.5\pm0.3\times 10^{-13}$} & \multirow{2}{*}{ $5.8\pm0.3\times 10^{-5}$} \\[-2pt]
2026-04-20 02:43:29 & 40.9 & Chandra/ACIS & 5.0 &  &  \\
2026-04-21 04:38:25 & 42.0 & EP/FXT-A+B & 3.2 &  $5.6\pm0.5\times 10^{-13}$ &  $5.9\pm0.5\times 10^{-5}$ \\
2026-04-23 05:06:26 & 44.0 & EP/FXT-A+B & 2.9 &  $5.0\pm0.5\times 10^{-13}$ &  $5.2\pm0.5\times 10^{-5}$ \\
2026-04-25 04:59:20 & 46.0 & EP/FXT-A+B & 3.0 & \multirow{2}{*}{ $5.3\pm0.3\times 10^{-13}$} & \multirow{2}{*}{ $5.5\pm0.3\times 10^{-5}$} \\[-2pt]
2026-04-25 09:46:04 & 46.2 & EP/FXT-A+B & 9.2 &  &  \\
2026-04-26 05:11:21 & 47.0 & XMM/pn+MOS & $\sim$20 &  $4.2\pm0.1\times 10^{-13}$ &  $4.4\pm0.1\times 10^{-5}$ \\
2026-05-17 07:43:43 & 68.1 & Chandra/ACIS & 9.9 &  $1.8\pm0.2\times 10^{-13}$ &  $1.9_{-0.2}^{+0.3}\times 10^{-5}$\\
\hline
\multicolumn{6}{l}{\textbf{Optical/NIR}}\\
Date& $\Delta t$& Telescope & Exposure & Filter & Magnitude \\
(UT) & [d] & & [ks] &  &  (AB)\\
\hline
2026-03-13 00:02:18 & 2.80 & T90  & 0.3 & $R$ & $18.35 \pm 0.02$ \\
2026-03-13 00:07:34 & 2.80 & T90  & 0.3 & $V$ & $18.49 \pm 0.03$ \\
2026-03-13 00:45:52 & 2.83 & T150  & 9.3 & $R$ & $18.41 \pm 0.01$ \\
2026-03-13 01:32:07 & 2.85 & CAHA  & 0.03 & $i$ & $17.97 \pm 0.08$ \\
2026-03-13 01:56:45 & 2.87 & BOOTES  & 0.9 & $r$ & $18.25 \pm 0.04$ \\
2026-03-13 02:04:11 & 2.88 & BOOTES  & 1.0 & $i$ & $18.02 \pm 0.05$ \\
2026-03-13 02:11:39 & 2.89 & BOOTES  & 1.0 & $Z$ & $17.85 \pm 0.05$ \\
2026-03-13 02:33:57 & 2.90 & T90  & 0.9 & $B$ & $18.79 \pm 0.06$ \\
2026-03-13 04:33:57 & 2.98 & T90  & 0.6 & $R$ & $18.34 \pm 0.03$ \\
2026-03-13 04:39:12 & 2.99 & T90  & 0.6 & $V$ & $18.43 \pm 0.01$ \\
2026-03-13 10:04:20 & 3.21 & BOOTES  & 3.3 & $g$ & $18.72 \pm 0.02$ \\
2026-03-13 10:27:04 & 3.23 & BOOTES  & 3.5 & $Z$ & $17.95 \pm 0.04$ \\
2026-03-13 10:36:59 & 3.24 & BOOTES  & 3.2 & $r$ & $18.43 \pm 0.02$ \\
2026-03-13 10:43:41 & 3.24 & BOOTES  & 3.2 & $i$ & $18.10 \pm 0.03$ \\
2026-03-13 20:59:38 & 3.67 & T90  & 1.5 & $V$ & $18.59 \pm 0.01$ \\
2026-03-13 21:17:43 & 3.68 & T150  & 0.9 & $R$ & $18.50 \pm 0.01$ \\
2026-03-13 21:36:35 & 3.69 & T90  & 2.4 & $B$ & $18.89 \pm 0.02$ \\
2026-03-13 22:02:17 & 3.71 & SBT  & 29.2 & $i$ & $18.27 \pm 0.10$ \\
2026-03-13 22:34:40 & 3.73 & T90  & 1.2 & $V$ & $18.55 \pm 0.02$ \\
2026-03-13 22:44:27 & 3.74 & O2m  & 1.3 & $g$ & $18.76 \pm 0.03$ \\
2026-03-13 23:07:52 & 3.76 & SBT  & 37.1 & $g$ & $18.68 \pm 0.05$ \\
2026-03-14 00:49:56 & 3.83 & BOOTES  & 1.8 & $r$ & $18.49 \pm 0.06$ \\
2026-03-14 01:02:18 & 3.84 & BOOTES  & 2.0 & $i$ & $18.25 \pm 0.09$ \\
2026-03-14 01:17:50 & 3.85 & BOOTES  & 2.1 & $Z$ & $18.07 \pm 0.06$ \\
2026-03-14 02:23:36 & 3.89 & D50  & 13.6 & $r$ & $18.48 \pm 0.02$ \\
2026-03-14 02:29:58 & 3.90 & T150  & 2.7 & $R$ & $18.63 \pm 0.01$ \\
2026-03-14 11:22:37 & 4.27 & BOOTES  & 1.5 & $g$ & $18.98 \pm 0.04$ \\
2026-03-14 11:23:50 & 4.27 & BOOTES  & 0.7 & $i$ & $18.36 \pm 0.09$ \\
2026-03-14 11:30:51 & 4.27 & BOOTES  & 0.7 & $Z$ & $18.02 \pm 0.08$ \\
2026-03-14 11:42:53 & 4.28 & BOOTES  & 1.1 & $r$ & $18.66 \pm 0.04$ \\
2026-03-14 18:30:39 & 4.56 & SBT  & 3.4 & $r$ & $18.62 \pm 0.10$ \\
2026-03-15 07:54:14 & 5.12 & BOOTES  & 1.3 & $g$ & $18.95 \pm 0.03$ \\
2026-03-15 08:08:44 & 5.13 & BOOTES  & 1.5 & $r$ & $18.70 \pm 0.03$ \\
2026-03-15 08:15:30 & 5.14 & BOOTES  & 1.5 & $i$ & $18.39 \pm 0.10$ \\
2026-03-15 08:39:12 & 5.15 & BOOTES  & 1.4 & $Z$ & $18.09 \pm 0.06$ \\
2026-03-16 00:25:03 & 5.81 & BOOTES  & 1.2 & $r$ & $18.90 \pm 0.07$ \\
2026-03-16 00:49:32 & 5.83 & BOOTES  & 1.0 & $i$ & $18.58 \pm 0.09$ \\
2026-03-16 03:43:11 & 5.95 & BOOTES  & 0.7 & $Z$ & $18.22 \pm 0.07$ \\
2026-03-16 08:07:33 & 6.13 & LUCI  & 0.2 & $J$ & $17.94 \pm 0.01$ \\
2026-03-16 08:07:33 & 6.13 & LUCI  & 0.2 & $K$ & $17.27 \pm 0.01$ \\
2026-03-16 22:59:36 & 6.75 & T150  & 0.3 & $I$ & $18.82 \pm 0.01$ \\
2026-03-16 23:04:50 & 6.76 & T150  & 0.3 & $R$ & $18.88 \pm 0.01$ \\
2026-03-16 23:10:05 & 6.76 & T150  & 0.3 & $V$ & $18.93 \pm 0.01$ \\
2026-03-17 00:32:00 & 6.82 & D50  & 2.1 & $r$ & $18.76 \pm 0.04$ \\
2026-03-17 04:14:15 & 6.97 & T150  & 0.6 & $I$ & $18.86 \pm 0.03$ \\
2026-03-17 04:37:22 & 6.99 & T150  & 0.6 & $R$ & $18.91 \pm 0.03$ \\
2026-03-17 04:47:41 & 6.99 & T150  & 0.6 & $V$ & $18.95 \pm 0.01$ \\
2026-03-17 05:44:39 & 7.03 & BOOTES  & 1.0 & $r$ & $18.90 \pm 0.03$ \\
2026-03-17 05:54:10 & 7.04 & BOOTES  & 1.0 & $i$ & $18.64 \pm 0.04$ \\
2026-03-17 06:03:42 & 7.05 & BOOTES  & 1.0 & $Z$ & $18.38 \pm 0.07$ \\
2026-03-17 07:45:04 & 7.12 & BOOTES  & 0.7 & $g$ & $19.17 \pm 0.03$ \\
2026-03-18 02:40:38 & 7.91 & D50  & 10.7 & $r$ & $18.84 \pm 0.04$ \\
2026-03-18 09:25:02 & 8.20 & LBC  & 0.1 & $z$ & $18.54 \pm 0.05$ \\
2026-03-18 09:25:07 & 8.20 & LBC  & 0.1 & $r$ & $18.84 \pm 0.08$ \\
2026-03-18 10:21:43 & 8.23 & BOOTES  & 0.7 & $r$ & $19.01 \pm 0.03$ \\
2026-03-18 10:25:38 & 8.23 & BOOTES  & 0.7 & $i$ & $18.78 \pm 0.04$ \\
2026-03-18 10:29:31 & 8.23 & BOOTES  & 0.7 & $Z$ & $18.58 \pm 0.07$ \\
2026-03-18 10:29:52 & 8.23 & BOOTES  & 1.0 & $g$ & $19.35 \pm 0.03$ \\
2026-03-19 02:06:51 & 8.88 & D50  & 14.4 & $r$ & $18.99 \pm 0.04$ \\
2026-03-20 02:52:01 & 9.91 & D50  & 8.7 & $r$ & $19.29 \pm 0.04$ \\
2026-03-22 00:12:28 & 11.80 & SBT  & 27.3 & $g$ & $19.49 \pm 0.11$ \\
2026-03-22 02:02:18 & 11.88 & D50  & 14.2 & $r$ & $19.44 \pm 0.03$ \\
2026-03-23 01:54:01 & 12.87 & D50  & 14.8 & $r$ & $19.55 \pm 0.04$ \\
2026-03-23 20:16:56 & 13.64 & BOOTES  & 0.9 & $i$ & $19.44 \pm 0.13$ \\
2026-03-24 01:46:40 & 13.87 & D50  & 14.2 & $r$ & $19.60 \pm 0.04$ \\
2026-03-25 07:53:04 & 15.12 & BOOTES  & 1.1 & $g$ & $20.17 \pm 0.08$\\
\hline
\multicolumn{6}{l}{Supplementary Optical/NIR data from public resources}\\
&0.012 & GOTO & &$L$ & $18.84 \pm 0.06$$^1$ \\
&0.100 & DDOTI & &$w$ & $16.60 \pm 0.30$$^2$ \\
&0.297 & ATLAS & &$orange$ & $16.65 \pm 0.03$$^3$ \\
&0.585 & LAST & &$Clear$ & $17.18 \pm 0.04$$^4$ \\
&0.590 & LAST & &$Clear$ & $17.17 \pm 0.04$$^4$ \\
&0.596 & LAST & &$Clear$ & $17.09 \pm 0.04$$^4$ \\
&0.600 & LAST & &$Clear$ & $17.16 \pm 0.04$$^4$ \\
&0.606 & LAST & &$Clear$ & $17.10 \pm 0.04$$^4$ \\
&0.611 & LAST & &$Clear$ & $17.11 \pm 0.04$$^4$ \\
&0.749 & LAST & &$Clear$ & $17.23 \pm 0.04$$^4$ \\
&0.792 & LAST & &$Clear$ & $17.25 \pm 0.04$$^4$ \\
&0.835 & LAST & &$Clear$ & $17.19 \pm 0.04$$^4$ \\
&0.879 & LAST & &$Clear$ & $17.30 \pm 0.04$$^4$ \\
&0.917 & LAST & &$Clear$ & $17.32 \pm 0.04$$^4$ \\
&1.12 & DDOTI & &$w$ & $17.52 \pm 0.01$$^2$ \\
&2.21 & WINTER & &$J$ & $17.16 \pm 0.09$$^5$ \\
&2.23 & WINTER & &$J$ & $17.28 \pm 0.09$$^5$ \\
&2.99 & WINTER & &$J$ & $17.34 \pm 0.12$$^5$ \\
&3.38 & SYSU & &$J$ & $17.59 \pm 0.10$$^6$ \\
&4.02 & TNG & &$J$ & $17.89 \pm 0.10$$^7$\\
\hline
\multicolumn{6}{l}{\textbf{Radio}}\\
Date & $\Delta t$ & Telescope & Exposure & Frequency & Flux density \\
(UT) & [d] & & [ks] & [GHz] & [mJy] \\
\hline
2026-03-13 23:20    & 3.7   & AMI-LA  &  28.8   & 15.2 (Ku)  & $9.8 \pm 0.5$$^8$  \\
2026-03-14 10:22    & 4.26  & VLA     & 0.898   & 6 (C)      & $3.9 \pm 0.2$$^9$  \\
2026-03-14 10:22    & 4.26  & VLA     & 0.778   & 10 (X)     & $6.4 \pm 0.3$$^9$  \\
2026-03-14 10:22    & 4.26  & VLA     & 0.539   & 15 (Ku)    & $9.1 \pm 0.5$$^9$  \\
2026-03-18 10:35    & 8.19  & VLA     & 0.898   & 6 (C)      & $7.1 \pm 0.4$ \\
2026-03-20          & 10    & NOEMA   &...      & 100 (W)    & $17.0 \pm 0.9$$^{10}$    \\
2026-03-26          & 16    & SMA     &...      & 230        & $5.7 \pm 0.2$$^{11}$  \\
2026-03-27 10:20    & 17.2  & VLA     & 0.359   & 3 (S)      & $3.04 \pm 0.04$   \\
2026-03-27 10:20    & 17.2  & VLA     & 0.479   & 6 (C)      & $7.82 \pm 0.06$  \\
2026-03-27 10:20    & 17.2  & VLA     & 0.479   & 10 (X)     & $11.01 \pm 0.07$  \\
2026-03-27 10:20    & 17.2  & VLA     & 0.598   & 15 (Ku)    & $11.69 \pm 0.11$  \\
2026-03-27 10:20    & 17.2  & VLA     & 1.197   & 33 (Ka)    & $9.50 \pm 0.24$   \\
2026-03-27 10:20    & 17.2  & VLA     & 1.795   & 45 (Q)     & $8.04 \pm 0.18$   \\
2026-04-04 10:30    & 25.2  & VLA     & 0.359   & 1.5 (L)    & $1.71 \pm 0.10$   \\
2026-04-04 10:30    & 25.2  & VLA     & 0.359   & 3 (S)      & $5.03 \pm 0.04$   \\
2026-04-04 10:30    & 25.2  & VLA     & 0.479   & 6 (C)      & $8.25 \pm 0.04$   \\
2026-04-04 10:30    & 25.2  & VLA     & 0.479   & 10 (X)     & $8.67 \pm 0.07$   \\
2026-04-04 10:30    & 25.2  & VLA     & 0.479   & 15 (Ku)    & $7.90 \pm 0.14$   \\
2026-04-04 10:30    & 25.2  & VLA     & 1.277   & 22 (K)     & $7.05 \pm 0.13$   \\
2026-04-04 10:30    & 25.2  & VLA     & 1.096   & 33 (Ka)    & $5.85 \pm 0.12$   \\
2026-04-04 10:30    & 25.2  & VLA     & 1.496   & 45 (Q)     & $4.93 \pm 0.08$   \\
2026-04-15 05:30    & 36.0  & VLA     &  0.363  & 1.5 (L)    &  $1.11 \pm 0.08$ \\
2026-04-15 05:30    & 36.0  & VLA     &  0.359  & 3 (S)      &  $5.31 \pm 0.05$ \\
2026-04-15 05:30    & 36.0  & VLA     &  0.479  & 6 (C)      &  $6.84 \pm 0.03$ \\
2026-04-15 05:30    & 36.0  & VLA     &  0.479  & 10 (X)     &  $6.55 \pm 0.11$ \\
2026-04-15 05:30    & 36.0  & VLA     &  0.455  & 15 (Ku)    &  $5.73 \pm 0.12$ \\
2026-04-15 05:30    & 36.0  & VLA     &  1.117  & 22 (K)     &  $4.83 \pm 0.10$ \\
2026-04-15 05:30    & 36.0  & VLA     &  0.987  & 33 (Ka)    &  $3.67 \pm 0.13$ \\
2026-04-15 05:30    & 36.0  & VLA     &  1.316  & 45 (Q)     &  $2.97 \pm 0.10$ \\
2026-04-29 08:40    & 50.2  & VLA     & 0.978   & 1.5 (L)    & $2.98 \pm 0.05$   \\
2026-04-29 08:40    & 50.2  & VLA     & 0.598   & 3   (S)    & $5.10 \pm 0.07$   \\
2026-04-29 08:40    & 50.2  & VLA     & 0.479   & 6   (C)    & $5.37 \pm 0.06$   \\
2026-04-29 08:40    & 50.2  & VLA     & 0.479   & 10  (X)    & $4.90 \pm 0.05$  \\ 
2026-05-05 04:25    & 55.0  & VLA     & 0.239   & 6  (C)     & $4.94 \pm 0.07$  \\
2026-05-05 04:25    & 55.0  & VLA     & 0.239   & 10 (X)     & $4.19 \pm 0.03$  \\
2026-05-05~04:25    & 55.0  & VLA     & 0.239   & 15 (Ku)    & $3.57 \pm 0.09$ \\
2026-05-11 07:20    & 62.1  & VLA     & 0.479   & 1.5 (L)    &   $2.86 \pm  0.07$                \\
2026-05-11 07:20    & 62.1  & VLA     & 0.359   & 3   (S)    &   $4.06 \pm  0.03$                \\
2026-05-11 07:20    & 62.1  & VLA     & 0.479   & 6   (C)    &   $4.31 \pm  0.05$                \\
2026-05-11 07:20    & 62.1  & VLA     & 0.479   & 10  (X)    &   $3.79 \pm  0.05$                \\
2026-05-11 07:20    & 62.1  & VLA     & 0.455   & 15  (Ku)   &   $3.08 \pm  0.04$                \\
2026-05-11 07:20    & 62.1  & VLA     & 1.117   & 22  (K)    &   ~$2.58 \pm  0.04$          
\end{longtable}
\end{ThreePartTable}

\bibliography{main}{}
\bibliographystyle{aasjournalv7}

\end{document}